\documentstyle[galley,epsfig]{mn2e}
\title{LAMOST carbon star}
\title[LAMOST carbon star \thanks]
{HCT/HESP study of two carbon stars from the LAMOST survey \thanks{Based on data collected using HCT/HESP}}
\author[J. Shejeelammal et al.]{J. Shejeelammal$^{1}$, Aruna Goswami$^{1}$, Jianrong Shi$^{2}$,  \\
$^{1}$Indian Institute of Astrophysics, Koramangala, Bangalore 560034,
India;  aruna@iiap.res.in\\ 
$^{2}$ CAS Key Laboratory of Optical Astronomy, National Astronomical Observatories, Beijing 100101, China. \\
}

\begin{document}

\date{ Accepted ---;  Received ---;  in original
form --- \large \bf }

\pagerange{\pageref{firstpage}--\pageref{lastpage}} \pubyear{2014}

\maketitle

 \begin{abstract}
Carbon stars, enhanced in carbon and neutron-capture elements, provide
wealth of information about the nucleosynthesis history of the Galaxy. 
In this work, we present the first ever detailed abundance analysis of
carbon star LAMOSTJ091608.81+230734.6 and a detailed abundance analysis of
neutron-capture elements for the object LAMOSTJ151003.74+305407.3.
Updates on the abundances of elements C, O, Mg, Ca, Cr, Mn and Ni for 
LAMOSTJ151003.74+305407.3 are also presented.
Our analysis is based on high resolution 
spectra obtained using Hanle  Echelle Spectrograph (HESP) attached to the 
Himalayan Chandra Telescope (HCT), IAO, Hanle. The stellar atmospheric 
parameters (T$_{eff}$, logg, micro-turbulance ${\zeta}$, metallicity [Fe/H])
are found to be (4820, 1.43, 1.62, $-$0.89) and (4500, 1.55, 1.24, $-$1.57)
for these two objects respectively. The abundance estimates of several 
elements, C, N, O, Na, $\alpha$-elements, Fe-peak elements and neutron-capture 
elements Rb, Sr, Y, Zr, Ba, La, Ce, Pr, Nd, Sm and Eu are presented. Our analysis 
shows the star LAMOSTJ151003.74+305407.3 to be a CEMP-r/s star, and 
LAMOSTJ091608.81+230734.6 a CH giant. We have examined if the i-process
model yields ([X/Fe]) of heavy elements could explain the observed 
abundances of the CEMP-r/s star based on a parametric model based analysis.
The negative values obtained for the neutron density dependent [Rb/Zr] 
ratio confirm  former  low-mass  
AGB companions for both the stars. Kinematic analysis shows 
that LAMOSTJ151003.74+305407.3 belongs to the Galactic halo population and 
LAMOSTJ091608.81+230734.6 to the disc population.
\end{abstract} 


\section{Introduction} 
Allowing to measure carbon and neutron-capture elements, the atmospheres
of the less-evolved low-mass stars form a unique treasure trove of information  
for the astrophysicists seeking the chemical evolution history of the Galaxy. 
Thus, studies on the metal-poor stars such as CH stars (Keenan 1942) with their more metal-poor
counterparts, Carbon Enhanced Metal-Poor (CEMP) stars offer the best means to 
constrain the neutron-capture nucleosynthesis processes, especially the nucleosythesis
occurring in the Asymptotic Giant Branch (AGB) stars. The spectra of these peculiar
stars show strong CH and C$_{2}$ molecular bands and features due to enhanced neutron-capture 
elements compared to the normal stars. They are characterized by C/O$>$ 1.  

\par The CEMP stars are more metal-poor ([Fe/H]$<$$-$1) than the classical CH stars 
(Lucatello et al. 2005, Aoki et al. 2007, Abate et al 2016, Hansen et al. 2016a, c)
with [C/Fe]$>$1 (Beers \& Christlieb 2005, Abate et al. 2016). They were first 
identified
among the Very Metal-Poor stars discovered in the extensive spectroscopic survey to 
identify a large sample of most metal-poor stars, HK survey (Beers et al. 1985, 1992, 2007
Beers 1999), and later in a number of successive surveys like Hamburg/ESO
Survey (HES; Christlieb et al. 2001a, 2001b, Christlieb 2003), Sloan Digital Sky Survey (SDSS; 
York et al. 2000), Sloan Extension for Galactic Understanding and Exploration (SEGUE; Yanny et al. 2009) etc.
A number of other  large sky survey programs in the past were also  
dedicated  to  
identify the Galactic carbon stars, for instance, the First
Byurakan Spectral Sky Survey (Gigoyan et al. 1998), 
the Automatic Plate Measuring survey (Totten \& Irwin 1998, Ibata et al. 2001), 
infrared objective-prism surveys (Alksnis et al. 2001),
Large Sky Area Multi-Object Fiber Spectroscopic Telescope (LAMOST) pilot
survey (Cui et al. 2012, Deng et al. 2012, Zhao et al. 2012).

Beers \& Christlieb (2005) put forward the first classification scheme
for CEMP stars and  classified them into different sub-classes, 
CEMP-s, CEMP-r, CEMP-r/s and CEMP-no depending 
on the level of enrichment of neutron-capture 
elements Ba and Eu.  A slight deviation from the original  classification 
schemes have been 
adopted by several authors (Aoki et al. 2007, Abate et al. 2016, Frebel 2018,
Hansen et al. 2019).  
High-resolution spectroscopic analyses have shown that, at present, 
about 80\% of the CEMP stars are CEMP-s stars (Aoki et al. 2007) and about 
half of the CEMP-s stars are CEMP-r/s stars 
(Sneden et al. 2008, K{\"a}ppeler et al. 2011, Bisterzo et al. 2011).

\par CH stars and CEMP-s/(r/s) stars belong to the main-sequence or giant phase of stellar 
evolution. Hence, the observed over abundances of the carbon and neutron-capture elements
are  attributed to an extrinsic origin. In the case of CH and CEMP-s stars, 
enriched in s-process elements, the most accepted scenario involves binary 
mass-transfer from an AGB companion. There exist a number of proposed 
scenarios for the simultaneous r- and s- process enrichment observed  in
CEMP-r/s stars (Jonsell et al. 2006 and references therein); however, none 
of them 
could successfully reproduce the observed frequency and high [hs/ls] ratio 
of CEMP-r/s stars (Abate et al. 2016). An intermediate neutron-capture 
process (i-process) that operates with neutron densities in  between s- and 
r-process neutron densities had  been invoked  to explain the
observed abundances of CEMP-r/s stars. Hampel et al. (2016, 2019) could 
successfully  reproduce the observed abundance trend of several CEMP-r/s stars
considering this production scenario. 
The i-process was originally proposed by Cowan \& Rose
(1977). 
Among the proposed scenarios for the nucleosynthetic sites of
the i-process are, massive (5 - 10 M$_{\odot}$) super-AGB stars 
(Doherty et al. 2015; Jones et al. 2016), evolved low-mass stars 
(Herwig et al. 2011; Hampel et al. 2019),  low-mass, low-metallicity 
([Fe/H] ${\le}$ $-$3) stars (Campbell \& Lattanzio 2008;
Campbell et al. 2010; Cruz et al. 2013; Cristallo et al. 2016), and 
Rapidly Accreting White Dwarfs (Herwig et al. 2014;
Denissenkov et al. 2017). Clarkson et al. (2018) and Banerjee et al. (2018)
have suggested that massive (M ${\ge}$  20 M$_{\odot}$), metal-poor
stars could also play a role in the production of i-process elements.
In spite of several efforts, large uncertainties still exist regarding 
the i-process nucleosynthesis and the possible astrophysical sites of 
its occurrence.(Frebel 2018; Koch et al. 2019).

It has been found from the long-term radial velocity 
monitoring studies that vast majority of CH stars (McClure et al. 1980, McClure 1983, 1984, McClure \&
Woodsworth 1990, Jorissen et al. 2016) and CEMP-s/rs stars (Lucatello et al. 2005, 
Starkenburg et al. 2014, Jorissen et al. 2016, Hansen et al. 2016c) are most likely binaries, 
thus strongly favoring the binary mass transfer scenario.  

\par It has been identified that the fraction of CEMP stars in the Galactic halo increases with decreasing metallicity; 
$\sim$20\% for [Fe/H]$<$$-$2 (Norris et al. 1997, Rossi et al. 1999, 2005, Christlieb 2003, Cohen et al. 2005,   
Marsteller et al. 2005, Frebel et al. 2006, Lucatello et al. 2006, Carollo et al. 2012, Lee et al. 2013), 
$\sim$40\% for [Fe/H]$<$$-$3 (Aoki et al. 2013, Lee et al. 2013, Yong et al. 2013b),
$\sim$75\% for [Fe/H]$<$$-$4 (Lee et al. 2013, Placco et al. 2014, Frebel \& Norris 2015), 
and thus making them important tools for the studies of formation and evolution of early Galactic halo.

Ji et al. (2016) have identified 
894 new carbon stars from the LAMOST DR2 which contains 
almost four million medium-resolution (R$\sim$1800)
stellar spectra, based on multiple line indices measurement. In this work, we 
have carried out a detailed spectroscopic analysis of two 
carbon stars LAMOSTJ091608.81+230734.6 and  
LAMOSTJ151003.74+305407.3 from  Ji et al. (2016).

\par Observations  and data reduction 
are presented  in Section 2.  
Radial velocity of the stars and the methodology used for the determination 
of stellar atmospheric parameters are presented in section 3. 
The same section also
provides a brief discussion on the stellar mass determination.  
Section 4 provides a discussion on abundance uncertainties. Elemental 
abundance determination is discussed   in Section 5. 
In Section 6, we present the kinematic analysis of the program stars, followed 
by the discussion on the binary status of the stars in Section 7. 
Interpretations of abundance ratios are presented in Section 8.
A discussion on individual stars along with the parametric model based 
analysis is 
also given in  section 8. Conclusion are drawn in Section 9.   
 
\section{OBSERVATIONS AND DATA REDUCTION} 
\begin{figure}
\centering
\includegraphics[width=\columnwidth]{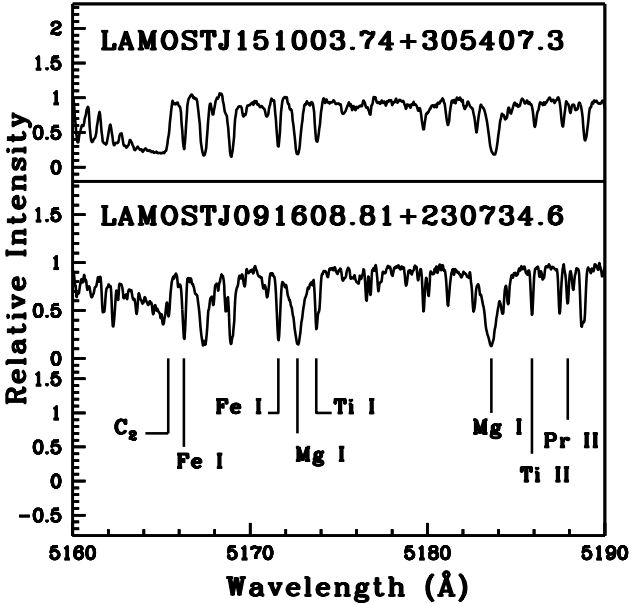}
\caption{ Sample spectra of the program stars in the  wavelength region 
5160 to 5190 {\bf  {\rm \AA}}.}\label{sample_spectra}
\end{figure}

High-quality, high-resolution spectra of the objects 
LAMOSTJ091608.81+230734.6 and LAMOSTJ151003.74+305407.3
were obtained with the  HESP (Hanle Echelle 
SPectrograph) attached to the 2m Himalayan Chandra Telescope 
(HCT) operated by Indian Astronomical  Observatory, Hanle.
The HESP spectra covers the wavelength range 3530 - 9970 {\rm \AA}. 
 The spectra of LAMOSTJ091608.81+230734.6  (V$_{mag}$ = 10.4) were 
 obtained on April 4, 2018  at a spectral resolution 
 ($\lambda/\delta\lambda$) $\sim$ 60,000, and the
spectra of LAMOSTJ151003.74+305407.3  (V$_{mag}$ = 11.38) were 
 obtained on May 23, 2018
at a spectral resolution ($\lambda/\delta\lambda$) $\sim$ 30,000.
For both  the objects we had acquired three frames; each frame 
was   taken with  2700 seconds exposure time. The three frames were 
added to enhance the S/N ratio, 
and the co-added spectrum was used for further analysis.
The data was reduced using the 
standard procedures in IRAF\footnote{IRAF (Image Reduction and Analysis Facility) 
is distributed by the National Optical Astronomical 
Observatories, which is operated by the Association for Universities 
for Research in Astronomy, Inc., under contract to the National 
Science Foundation} software. The basic information of the program stars are given in
Table \ref{basic data of program stars}. Two sample spectra of the stars in the 
wavelength region 5160 - 5190 {\rm \AA} are shown in Figure \ref{sample_spectra}.\\

 {\footnotesize
\begin{table*}
\caption{\textbf{Basic information of the program stars.}\label{basic data of program stars}}
\resizebox{\textwidth}{!}{
\begin{tabular}{lccccccccccc}
\hline
Star                        & RA$(2000)$      & Dec.$(2000)$    & B       & V       & J        & H        & K      & Exposure       & Date of obs.   & S/N      &  \\
                            &                 &                 &         &         &          &          &        & (seconds)      &                & 5500 \AA & 7500 \AA \\
\hline
LAMOSTJ091608.81+230734.6  & 09 16 8.82      & +23 07 34.86    & 11.44  & 10.40    & 8.654    & 8.141    & 8.022  & 2700(3)        & 04/04/2018      & 37.15    & 42.38  \\      
LAMOSTJ151003.74+305407.3  & 15 10 3.30      & +30 54 7.36     & 13.50  & 11.38    & 9.33     & 8.737    & 8.539  & 2700(3)        & 23/05/2018      & 33.67    & 70.17  \\
\hline
\end{tabular}}

The number of frames taken are indicated within the parenthesis with exposure.
\end{table*} 
}
 
\section{ESTIMATION OF ATMOSPHERIC PARAMETERS AND RADIAL VELOCITY}
A set of clean lines of several elements is used to calculate the 
radial velocities of the program stars.
While LAMOSTJ151003.74+305407.3 is found to be a high velocity
object with an estimated  radial velocity of $-$141.58$\pm$3.57 kms$^{-1}$,
the radial velocity of LAMOSTJ091608.81+230734.6 is found to be 
16.13$\pm$4.30 kms$^{-1}$. The radial velocities of these two objects 
are $-$145.2 and 17.3 kms$^{-1}$ respectively, as noted from 
the SIMBAD astronomical database (Gaia collaboration et al. 2018). 
For the star LAMOSTJ151003.74+305407.3, our estimate shows a difference of $\sim$4 kms$^{-1}$
from the SIMBAD value; this may be a clear indication that the star could be a binary.  

The equivalent width measurements of a set of Fe I and Fe II lines
are used to derive the atmospheric parameters of the stars. The lines are selected
such that the range of equivalent width and excitation potential are
20 - 180 {\rm m\AA} and 0.0 - 6.0 eV respectively. 
The photometric temperature estimates, estimated 
using the temperature calibration equations of Alonso et al. (1999, 2001), 
had been used 
as an initial guess to derive the stellar atmospheric parameters. 
The final model atmosphere is obtained through an iterative process from the 
initial one taken from the Kurucz grid of model atmosphere 
with no convective overshooting (http://cfaku5.cfa.hardvard.edu/). 
We made use of the recent  version MOOG2013
of the radiative transfer code MOOG (Sneden 1973) assuming
Local Thermodynamic Equilibrium (LTE) for the analysis. 

\begin{figure}
\centering
\includegraphics[width=\columnwidth]{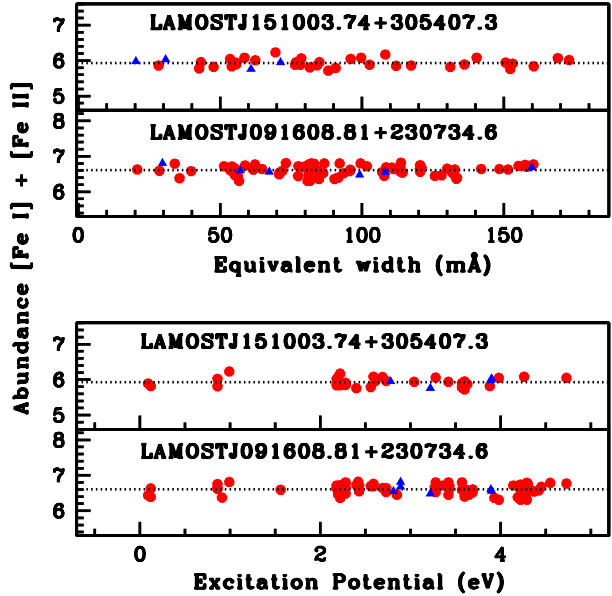}
\caption{Iron abundances of the program stars derived from individual Fe I and Fe II 
lines as function of (i) Excitation potential (lower panel),
(ii) equivalent width (upper panel). 
Solid circles correspond to Fe I and solid 
triangles correspond to Fe II lines. } \label{ep_ew}
\end{figure}

The temperature which gives nearly zero slope between the 
abundance and excitation potential of Fe I lines is taken as the 
effective temperature. The microturbulent velocity at this fixed
effective temperature is then determined such that there is no dependence
of the Fe I abundance on equivalent width. With this temperature and
microturbulent velocity estimates, the surface gravity is determined by
demanding the abundances obtained from Fe I and Fe II lines are to be nearly same.
Figure \ref{ep_ew} shows the abundances estimated from Fe I and Fe II lines, 
as functions of excitation potential and equivalent widths.
The derived atmospheric parameters and the radial velocity
estimates are given in Table \ref{atmospheric parameters}. 

{\footnotesize
 \begin{table*}
\caption{Derived atmospheric parameters of the program stars.} \label{atmospheric parameters}
\begin{tabular}{lcccccccccc}
\hline
Star                            & T$\rm_{eff}$  & log g            & $\zeta$           & [Fe I/H]          &[Fe II/H]            & V$_{r}$             &  V$_{r}$            & Remarks\\
                                &    (K)        & cgs              & (km s$^{-1}$)     &                   &                     & (km s$^{-1}$)       & (km s$^{-1}$)       &   \\
                                & $\pm$100      & $\pm$0.2         & $\pm$0.2          &                   &                     & our estimates      &  SIMBAD              &   \\  
\hline
LAMOSTJ091608.81+230734.6      & 4820          & 1.43             & 1.62              & $-$0.89$\pm$0.14  & $-$0.89$\pm$0.12    & +16.13$\pm$4.30     & +17.33$\pm$0.40     & 1  \\
LAMOSTJ151003.74+305407.3      & 4500          & 1.55             & 1.24              & $-$1.57$\pm$0.12  & $-$1.57$\pm$0.12    & $-$141.58$\pm$3.57  & $-$145.25$\pm$0.003 & 1  \\
                                & 4358.31       & 0.956            & 1.667             & $-$1.346          & --                  & --                  & --                  & 2  \\
\hline
\end{tabular}

Remarks: 1. Our work, 2. Hayes et al. (2018) \\
\end{table*}
}

\par
\begin{figure}
\centering
\includegraphics[width=\columnwidth]{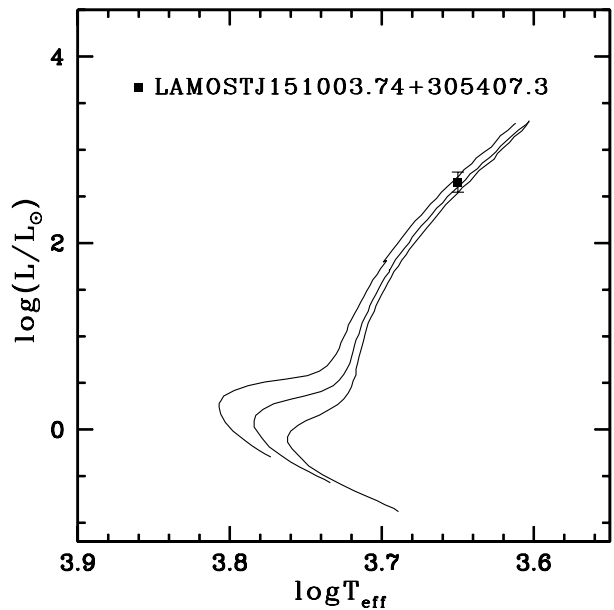}
\caption{The locations of LAMOSTJ151003.74+305407.3. 
The evolutionary tracks for 0.6, 0.7 and 0.8 M$_{\odot}$ 
are shown from bottom to top for z = 0.001.} \label{track_001}
\end{figure}

 We have determined the mass of the star LAMOSTJ151003.74+305407.3 
from its position in the Hertzsprung-Russell diagram, generated using the 
evolutionary tracks of Girardi et al. 2000, with the estimates of spectroscopic 
temperature, T$\rm_{eff}$, and the luminosity, log$(L/L_{\odot})$.
Then log\,{g} is recalculated using this mass estimate as described 
in our previous work Shejeelammal et al. (2020). For estimation 
of log$(L/L_{\odot})$, the required 
 visual magnitudes V and the parallaxes $\pi$ are taken from Simbad 
and Gaia DR2 (Gaia collaboration et al. 2018, https://gea.esac.esa.int/archive/) respectively. We have used 
z = 0.001 tracks for this star.
The evolutionary tracks for LAMOSTJ151003.74+305407.3 is shown in 
Figure \ref{track_001}.
The estimated mass, log\,{g} determined using parallax method  
are presented in Table \ref{mass age}.
The mass could not be
determined for the other star using this method as the evolutionary 
tracks corresponding 
to its temperature and luminosity are not available.

{\footnotesize
\begin{table*}
\caption{Mass and log\,{g} estimates by parallax method} \label{mass age}
\begin{tabular}{lcccccc}
\hline                       
 Star name                        & Parallax             & $M_{bol}$        & log(L/L$_{\odot}$)  & Mass(M$_{\odot}$) & log g          & log g (spectroscopic)   \\
                                  & (mas)                &                  &                     &                   & (cgs)          & (cgs)                    \\
\hline
LAMOSTJ091608.81+230734.6        & 1.0543$\pm$0.0517    & 0.105$\pm$0.107  & 1.854$\pm$0.043     & --                & --             & 1.43                      \\
LAMOSTJ151003.74+305407.3        & 0.2778$\pm$0.0346    & 1.891$\pm$0.271  & 2.653$\pm$0.109     & 0.70$\pm$0.10     & 1.20$\pm$0.05  & 1.55                   \\
\hline
\end{tabular}
\end{table*}
}

\section{ABUNDANCE UNCERTAINTIES}
The main sources of uncertainty in the abundance are the errors in the 
estimated stellar atmospheric parameters and the errors in the line parameters.
Total uncertainty in the elemental abundance, log $\epsilon$, is 
given as; \\

$\sigma_{log\epsilon}^{2}$ = $\sigma_{ran}^{2}$ + $(\frac{\partial log \epsilon}{\partial T})^{2}$ $\sigma_{T_{eff}}^{2}$ + $(\frac{\partial log \epsilon}{\partial log g})^{2}$ $\sigma_{log g}^{2}$ + \\
\begin{center}
  $(\frac{\partial log \epsilon}{\partial \zeta})^{2}$ $\sigma_{\zeta}^{2}$ + $(\frac{\partial log \epsilon}{\partial [Fe/H]})^{2}$ $\sigma_{[Fe/H]}^{2}$ \\
\end{center}

\noindent where $\sigma_{ran}$ = $\frac{\sigma_{s}}{\sqrt{N}}$, $\sigma_{s}$ 
being the standard deviation of the abundances derived from N 
lines of a particular element considered. The other $\sigma$'$_{s}$ in the equation are 
the typical uncertainties in the atmospheric parameters;   
$\Delta$T$_{eff}$$\sim$ $\pm$100 K, $\Delta$log g$\sim$ $\pm$0.2 dex, 
$\Delta$$\zeta$$\sim$ $\pm$0.2 km s$^{-1}$ and $\Delta$[Fe/H]$\sim$ $\pm$0.1 dex.

\par We have calculated the uncertainty in the abundance, [X/Fe], for each element
following a detailed procedure described in our earlier paper Shejeelammal et al. (2020).
As an example, these values for the star LAMOSTJ091608.81+230734.6 are given in Table \ref{differential_abundance}.

{\footnotesize
\begin{table*}
\caption{Differential Abundance ($\Delta$log$\epsilon$) of different elemental species due to the variations
in stellar atmospheric parameters for LAMOSTJ091608.81+230734.6. The sixth column gives 
the computed rms uncertainty of the second to fifth columns. The seventh column gives the 
total uncertainty in the abundance ratio, [X/Fe], of each elemental species.}  
\label{differential_abundance}
\resizebox{\textwidth}{!}{\begin{tabular}{lcccccc}
\hline                       
Element & $\Delta$T$_{eff}$  & $\Delta$log g  & $\Delta$$\zeta$       & $\Delta$[Fe/H] & ($\Sigma \sigma_{i}^{2}$)$^{1/2}$ & $\sigma_{[X/Fe]}$  \\
        & ($\pm$100 K)       & ($\pm$0.2 dex) & ($\pm$0.2 kms$^{-1}$) & ($\pm$0.1 dex) &                                   &       \\
\hline
C	        & $\pm$0.10	       & 0.00	          & 0.00	                & $\mp$0.05	     & 0.11	  & 0.19   \\
N	        & $\pm$0.13	       & $\pm$0.03	    & 0.00	                & $\pm$0.02	     & 0.13	  & 0.21   \\
O	        & 0.00	           & $\pm$0.04	    & $\mp$0.02	            & 0.00	         & 0.04	  & 0.16   \\
Na I	    & $\pm$0.08	       & $\mp$0.01	    & $\mp$0.05	            & $\mp$0.01	     & 0.10	  & 0.19   \\
Mg I	    & $\pm$0.09	       & $\mp$0.03	    & $\mp$0.08	            & $\mp$0.01	     & 0.05	  & 0.17   \\
Si I	    & $\pm$0.04	       & $\pm$0.01	    & $\mp$0.02	            & 0.00	         & 0.06	  & 0.17  \\
Ca I	    & $\pm$0.10	       & $\mp$0.01	    & $\mp$0.08	            & $\mp$0.01	     & 0.13	  & 0.21   \\
Sc II	    & $\mp$0.02	       & $\pm$0.09	    & $\mp$0.06	            & $\pm$0.02	     & 0.11	  & 0.25   \\
Ti I	    & $\pm$0.14	       & $\mp$0.01	    & $\mp$0.07	            & $\mp$0.01	     & 0.16	  & 0.23   \\
Ti II	    & 0.00	           & $\pm$0.09	    & $\mp$0.08	            & $\pm$0.03	     & 0.12	  & 0.26    \\
V I	      & $\pm$0.16	       & $\mp$0.01	    & $\mp$0.07	            & $\mp$0.01	     & 0.18	  & 0.24   \\
Cr I	    & $\pm$0.17	       & $\mp$0.02	    & $\mp$0.13	            & $\mp$0.03	     & 0.22	  & 0.27   \\
Mn I	    & $\pm$0.09	       & $\mp$0.02	    & $\mp$0.16	            & $\mp$0.01	     & 0.18	  & 0.24   \\
Fe I	    & $\pm$0.10	       & $\pm$0.01	    & $\mp$0.11	            & $\pm$0.05  	   & 0.16  	& -- \\
Fe II	    & $\mp$0.06	       & $\pm$0.11	    & $\mp$0.10	            & $\mp$0.17  	   & 0.23  	& -- \\
Co I	    & $\pm$0.07	       & $\pm$0.02	    & $\mp$0.06	            & $\pm$0.01	     & 0.09	  & 0.18   \\
Ni I	    & $\pm$0.09	       & 0.00	          & $\mp$0.05	            & $\mp$0.01	     & 0.10	  & 0.19   \\
Zn I	    & $\mp$0.03	       & $\pm$0.05	    & $\mp$0.03	            & $\pm$0.02	     & 0.07	  & 0.17   \\
Rb I	    & $\pm$0.10	       & 0.00	          & $\mp$0.03	            & 0.00	         & 0.10	  & 0.19   \\
Sr I	    & $\pm$0.18	       & $\mp$0.02	    & $\mp$0.13	            & $\mp$0.02	     & 0.22	  & 0.27   \\
Y I	      & $\pm$0.17	       & $\mp$0.02	    & $\mp$0.04	            & $\mp$0.02	     & 0.18	  & 0.24   \\
Y II	    & 0.00	           & $\pm$0.09	    & $\mp$0.11	            & $\pm$0.03	     & 0.14	  & 0.27   \\
Zr I	    & $\pm$0.17	       & $\mp$0.02	    & $\mp$0.07	            & $\mp$0.02	     & 0.19	  & 0.25   \\
Zr II	    & $\pm$0.01	       & $\pm$0.08	    & $\mp$0.06	            & $\pm$0.02	     & 0.10	  & 0.25   \\
Ba II	    & $\pm$0.02	       & $\pm$0.05	    & $\mp$0.15	            & $\pm$0.02	     & 0.16	  & 0.28   \\
La II     & $\pm$0.01        & $\pm$0.09      & $\mp$0.06             & $\pm$0.03      & 0.11   & 0.25    \\
Ce II	    & $\pm$0.02	       & $\pm$0.08	    & $\mp$0.13	            & $\pm$0.02	     & 0.16	  & 0.28   \\
Pr II	    & $\pm$0.02	       & $\pm$0.08	    & $\mp$0.07	            & $\pm$0.02	     & 0.11	  & 0.25   \\
Nd II	    & $\pm$0.03	       & $\pm$0.08	    & $\mp$0.16	            & $\pm$0.02	     & 0.18	  & 0.29   \\
Sm II	    & $\pm$0.04	       & $\pm$0.09	    & $\mp$0.08	            & $\pm$0.03	     & 0.13	  & 0.26   \\
Eu II	    & $\mp$0.02	       & $\pm$0.09	    & $\mp$0.03	            & $\pm$0.03	     & 0.10	  & 0.25   \\
\hline
\end{tabular}}

\end{table*}
}

\section{Abundance determination}
Elemental abundances are derived from the measured equivalent widths and
the spectral synthesis calculation of spectral lines of neutral and
ionized atoms. Only clean unblended lines have been considered for
the abundance determination. Absorption lines due to each element is identified
from the close comparison of Doppler-corrected spectrum of the star Arcturus and 
the program stars spectra. The line information such as
log $gf$ and the lower excitation potential are from the Kurucz
database of atomic line lists. In addition to the equivalent width method,
spectral synthesis calculation is also performed for the elements
showing hyper-fine splitting. Also, the abundances derived from the
molecular bands are based on spectral synthesis calculation.
The hyper-fine components of Eu are taken from Worely
et al. (2013), Ba from Mcwilliam (1998), V, Co and Cu from Prochaska et al. (2000)
and Sc and Mn from Prochaska \& Mcwilliam (2000).
Solar abundance values are taken from Asplund et al. (2009).

\par The abundance results are presented in Table \ref{abundance_table}.
The lines used for the estimation of abundances are given in
Tables A1 and A2.

{\footnotesize
\begin{table*}
\caption{Elemental abundances in LAMOSTJ091608.81+230734.6 and LAMOSTJ151003.74+305407.3} \label{abundance_table}
\resizebox{\textwidth}{!}
{\begin{tabular}{lcccccccccccc}
\hline
       &    &                              &                    & LAMOSTJ091608.81+230734.6 &            &                    & LAMOSTJ151003.74+305407.3 &         \\ 
\hline
       & Z  & solar log$\epsilon^{\ast}$   & log$\epsilon$      & [X/H]    & [X/Fe]     & log$\epsilon$      & [X/H]    & [X/Fe] \\ 
\hline 
C (C$_{2}$ band 5165 {\rm \AA})     & 6  & 8.43                         & 7.90(syn)          & $-$0.53  & 0.36       & 8.60(syn)          & 0.17     & 1.74  \\
C (C$_{2}$ band 5635 {\rm \AA})     & 6  & 8.43                         & 7.90(syn)          & $-$0.53  & 0.36       & 8.60(syn)          & 0.17     & 1.74  \\
N      & 7  & 7.83                         & 7.79(syn)          & $-$0.04  & 0.85       & 7.53(syn)          & $-$0.30  & 1.27     \\
O      & 8  & 8.69                         & 7.76(syn)          & $-$0.93  & $-$0.04    & --                 & --       & --     \\
Na I   & 11 & 6.24                         & 6.03$\pm$0.15(4)   & $-$0.21  & 0.68       & 5.24$\pm$0.05(2)   & $-$1.00  & 0.57      \\
Mg I   & 12 & 7.60                         & 7.12$\pm$0.15(2)   & $-$0.46  & 0.43       & 6.20$\pm$0.18(3)   & $-$1.58  & $-$0.01  \\
Si I   & 14 & 7.51                         & 6.59$\pm$0.18(3)   & $-$0.92  & $-$0.03    & --                 & --       & --        \\
Ca I   & 20 & 6.34                         & 5.38$\pm$0.15(10)  & $-$0.96  & $-$0.07    & 4.52$\pm$0.13(8)   & $-$1.82  & $-$0.25  \\
Sc II  & 21 & 3.15                         & 2.19(syn)          & $-$0.96  & $-$0.07    & 1.85(syn)          & $-$1.30  & 0.27     \\
Ti I   & 22 & 4.95                         & 4.07$\pm$0.17(12)  & $-$0.88  & 0.01       & 3.30$\pm$0.14(4)   & $-$1.65  & $-$0.08   \\
Ti II  & 22 & 4.95                         & 3.77$\pm$0.17(4)   & $-$1.18  & $-$0.29    & 3.21$\pm$0.19(5)   & $-$1.74  & $-$0.17     \\
V I    & 23 & 3.93                         & 3.42(syn)          & $-$0.51  & 0.38       & 3.11(syn)          & $-$0.82  & 0.75   \\
Cr I   & 24 & 5.64                         & 5.13$\pm$0.18(6)   & $-$0.51  & 0.38       & 3.59$\pm$0.10(5)   & $-$2.05  & $-$0.48  \\
Mn I   & 25 & 5.43                         & 4.70(syn)          & $-$0.73  & 0.16       & 3.43(syn)          & $-$2.00  & $-$0.43  \\
Fe I   & 26 & 7.50                         & 6.61$\pm$0.14(73)  & $-$0.89  & -          & 5.93$\pm$0.12(35)  & $-$1.57  & -          \\
Fe II  & 26 & 7.50                         & 6.61$\pm$0.12(6)   & $-$0.89  & -          & 5.93$\pm$0.12(4)   & $-$1.57  & -        \\
Co I   & 27 & 4.99                         & 3.89(syn)          & $-$1.10  & $-$0.21    & --                 & --       & --    \\
Ni I   & 28 & 6.22                         & 5.63$\pm$0.11(11)  & $-$0.59  & 0.30       & 4.61$\pm$0.17(6)   & $-$1.61  & $-$0.04     \\
Zn I   & 30 & 4.56                         & 4.11(1)            & $-$0.49  & 0.40       & 2.56$\pm$0.14(2)   & $-$2.00  & $-$0.43      \\
Rb I   & 37 & 2.52                         & 2.10(syn)          & $-$0.42  & 0.47       & 1.00(syn)          & $-$1.52  & 0.05   \\
Sr I   & 38 & 2.87                         & 2.95(syn)          & 0.08     & 0.97       & --                 & --       & --        \\
Y I    & 39 & 2.21                         & 2.56(syn)          & 0.35     & 1.24       & 2.20(syn)          & $-$0.01  & 1.56       \\
Y II   & 39 & 2.21                         & 2.32$\pm$0.10(6)   & 0.11     & 1.00       & 3.05$\pm$0.11(2)   & 0.84     & 2.40     \\
Zr I   & 40 & 2.58                         & 3.08(syn)          & 0.50     & 1.39       & 2.05(syn)          & $-$0.53  & 1.04     \\
Zr II  & 40 & 2.58                         & 2.73(syn)          & 0.15     & 1.04       & 2.09(syn)          & $-$0.49  & 1.08       \\
Ba II  & 56 & 2.18                         & 2.48(syn)          & 0.30     & 1.19       & 2.00(syn)          & $-$0.18  & 1.39     \\
La II  & 57 & 1.10                         & 2.00(syn)          & 0.90     & 1.79       & 1.10(syn)          & 0.00     & 1.57 \\ 
Ce II  & 58 & 1.58                         & 2.34$\pm$0.11(7)   & 0.76     & 1.65       & 1.32$\pm$0.13(7)   & $-$0.26  & 1.31     \\
Pr II  & 59 & 0.72                         & 1.48$\pm$0.18(6)   & 0.76     & 1.65       & 1.02$\pm$0.11(5)   & 0.30     & 1.87     \\
Nd II  & 60 & 1.42                         & 2.17$\pm$0.17(9)   & 0.75     & 1.64       & 1.35$\pm$0.14(12)  & $-$0.07  & 1.50     \\
Sm II  & 62 & 0.96                         & 1.41$\pm$0.11(10)  & 0.45     & 1.34       & 1.03$\pm$0.08(6)   & 0.07     & 1.64      \\
Eu II  & 63 & 0.52                         & 0.33(syn)          & $-$0.19  & 0.70       & 0.09(syn)          & $-$0.43  & 1.14     \\
\hline
\end{tabular}}

$\ast$  Asplund (2009), The numbers within the parenthesis are
the number of lines used for abundance estimation.
\end{table*}
}

\subsection{Abundance analysis: C, N, O, $^{12}$C/$^{13}$C, 
Na, $\alpha$- and $Fe$-peak elements}
The abundance of oxygen is derived using the spectral synthesis
calculation of [O I] line at 6300.304 {\rm \AA} in 
LAMOSTJ091608.81+230734.6. 
The resonance  O I triplet lines at around 7770 {\rm \AA} and the
[O I] 6363.776  {\rm \AA} line are blended and could not be 
used for abundance determination. 
The estimated oxygen abundance shows near solar value, with 
[O/Fe]$\sim$$-$0.04. We could not estimate the abundance of oxygen 
for the star LAMOSTJ151003.74+305407.3 as no clean good lines of
oxygen could be detected in the  spectrum of this object.

\par
\begin{figure}
\centering
\includegraphics[width=0.8\columnwidth, height= 0.8\columnwidth]{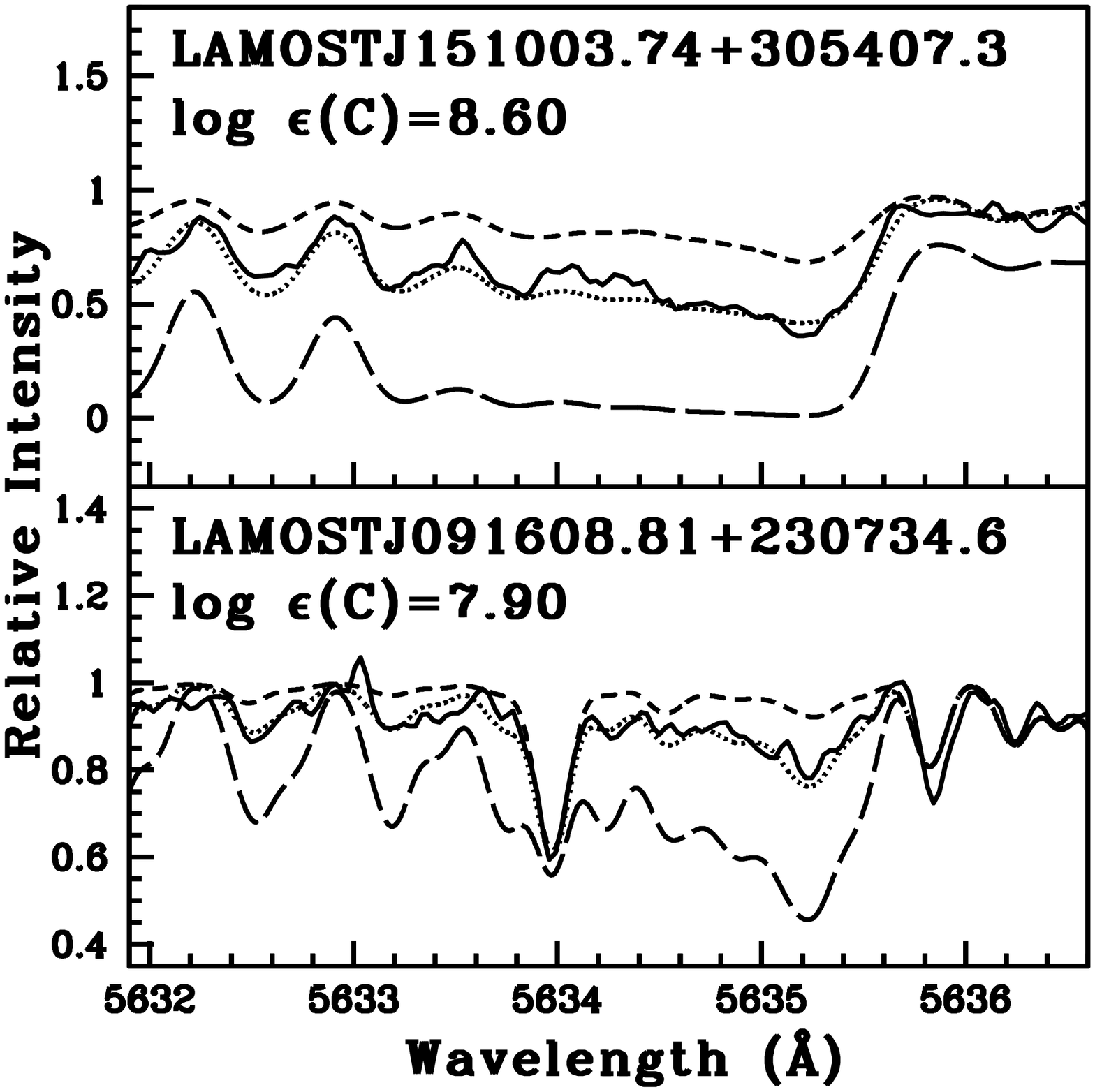}
\includegraphics[width=0.8\columnwidth, height= 0.8\columnwidth]{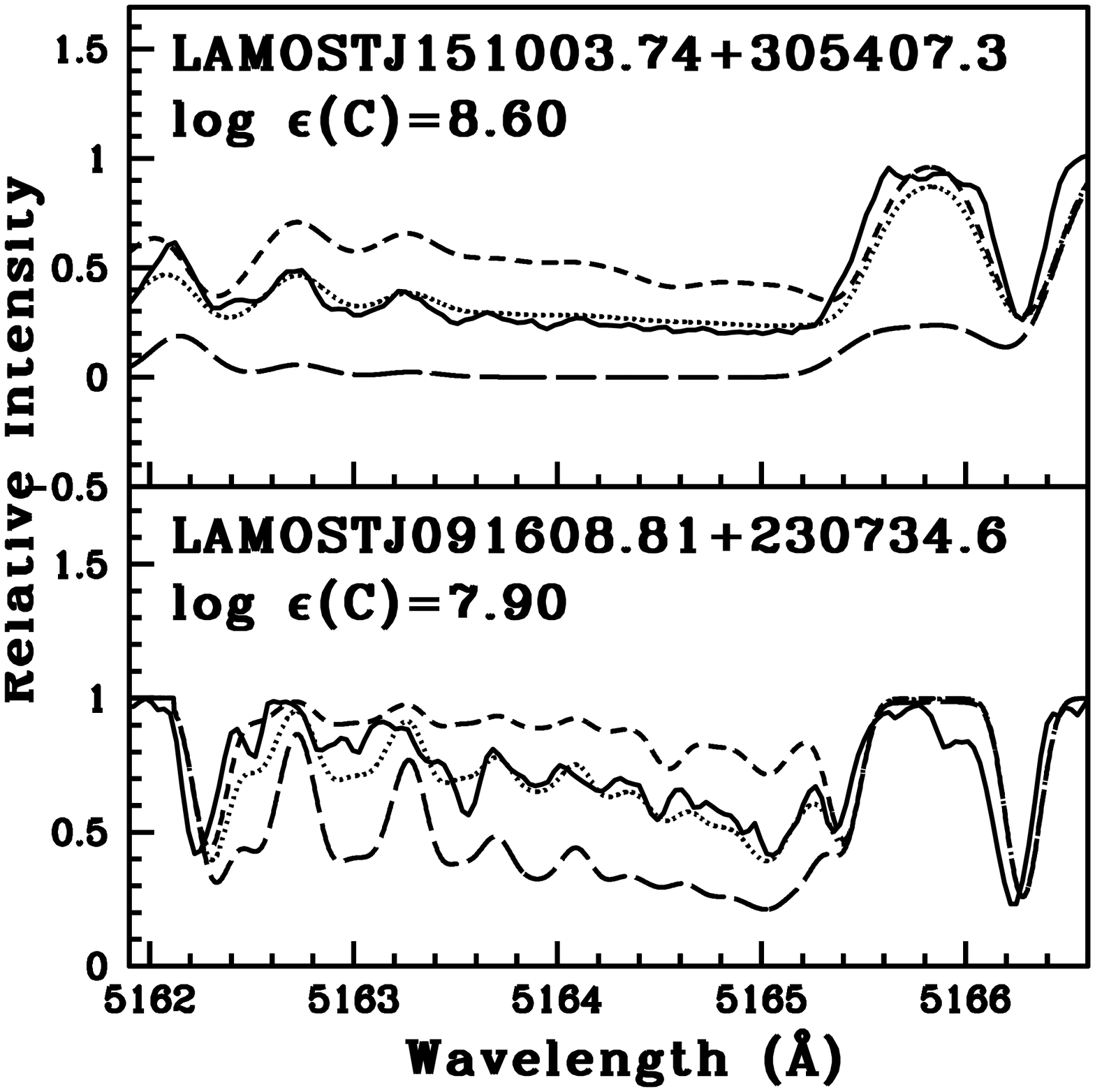}
\caption{ Synthesis of C$_{2}$ band around 5165 {\rm \AA} (lower panel) and 5635 {\rm \AA} (upper panel). 
Dotted and solid lines represent synthesized and observed spectra respectively. 
Short dashed and long dashed lines represent the synthetic spectra 
for $\Delta$ [C/Fe] = $-$0.3 and +0.3 respectively.} \label{carbon}
\end{figure}

 The carbon abundance is derived from the spectral synthesis 
calculation of the C$_{2}$ molecular bands at 5165 and 5635 {\rm \AA}
(Figure \ref{carbon}). Both the bands gave the same carbon abundance 
value in both the stars. 
Carbon is found to be enhanced in LAMOSTJ151003.74+305407.3 with 
[C/Fe]$\sim$1.74 and mildly enhanced in the other object with 
[C/Fe]$\sim$0.36.  

\par Once the carbon abundance is estimated, the abundance of nitrogen
is derived using the spectral synthesis calculation of $^{12}$CN lines at 8000 {\rm \AA} region.
The $^{12}$CN molecular band at 4215 {\rm \AA} is not usable in both the stars. 
The CN and C$_{2}$ molecular lines are taken from Ram et al. (2014), Sneden et al. (2014) 
and Brooke et al. (2013). In both the stars
nitrogen is enhanced with [N/Fe]$\geq$0.85. 
The final C, N and O abundances are determined by an iterative process.
Using the first estimate of oxygen abundance derived from the spectral synthesis
calculation of 6300.304 {\rm \AA} [O I] line, the abundance of carbon is estimated 
from the C$_{2}$ molecular bands at 5165 and 5635 {\rm \AA}. The abundance of 
nitrogen is then determined using these derived abundance estimates of O and C. 
Once the nitrogen abundance is obtained, the oxygen and carbon abundances are re-determined
using this nitrogen abundance. This iteration process is continued until 
a convergence is reached.

\begin{figure}
\centering
\includegraphics[width=\columnwidth]{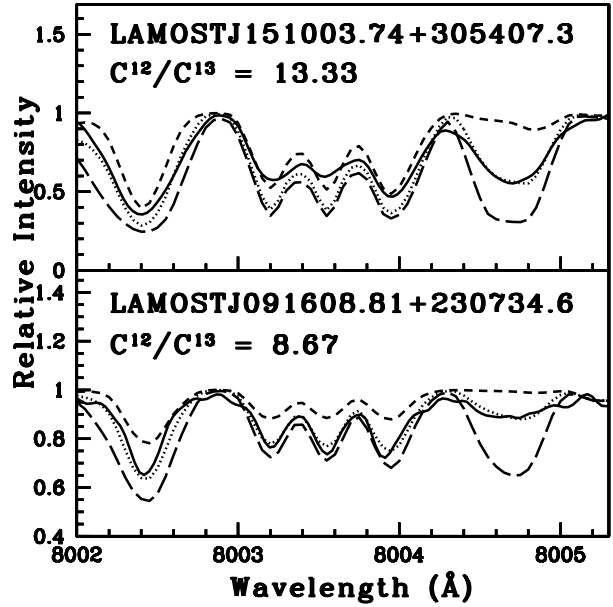}
\caption{Spectral synthesis of CN band around 8005 {\rm \AA}. 
Dotted and solid lines represent synthesized and observed spectra respectively.
Short dashed and long dashed lines are the synthetic spectra for  
$^{12}$C/$^{13}$C $\simeq$ 83 and 2.7 respectively.} \label{13c_band}
\end{figure}

The carbon isotopic ratio, $^{12}$C/$^{13}$C, is estimated
from the spectral synthesis calculation of $^{12}$CN lines at 
8003.292, 8003.553, 8003.910  {\rm \AA},  and $^{13}$CN features at 8004.554, 
8004.728, 8004.781 {\rm \AA}. The spectrum synthesis 
fits for the program stars in this region is shown in Figure \ref{13c_band}.
The values of this ratio are 8.67 and 13.33 
in LAMOSTJ091608.81+230734.6 and LAMOSTJ151003.74+305407.3 respectively. 
These are  typical values normally found in the case of  giants (Smith et al. 1993).
For CEMP-s and CEMP-r/s stars, the values of $^{12}$C/$^{13}$C ratio are
found in the range 2.5 - 40 (Bisterzo et al. 2011).
We could estimate the C/O ratio in the star LAMOSTJ091608.81+230734.6
which is found to be greater than 1 as typically seen in CH stars.

\par The abundances of the elements Na, Mg, Si, Ca, Ti, Cr, Ni and Zn are derived
from the measured equivalent width of spectral lines listed in Table A2.
Scandium abundance is derived from the spectral synthesis calculation of 
Sc II line at 6245.637 {\rm \AA}, vanadium abundance from the V I lines at 6251.827 and 
4864.731 {\rm \AA}, manganese abundance from Mn I line at 6013.513 {\rm \AA} and 
the abundance of cobalt from Co I line at  5342.695 {\rm \AA}. 

A comparison of our estimated light element abundances for the star LAMOSTJ151003.74+305407.3 
with the literature values is given in Table \ref{J151003_abundance_literature}. 
Within the error limits, our estimates of Mg and Ni match with the estimates of Hayes et al. (2018). 
However, our estimates are lower by $\sim$0.6 dex for [Ca/Fe] and [Cr/Fe] and $\sim$0.3 dex for [Mn/Fe]. 
We have obtained [C/Fe]$\sim$1.74, which is largely different from $\sim$0.81 of Hayes et al. (2018).

{\footnotesize
\begin{table*}
\caption{\textbf{Comparison of the light element abundances of LAMOSTJ151003.74+305407.3 with the literature values.}}  \label{J151003_abundance_literature}
\resizebox{\textwidth}{!}{\begin{tabular}{lcccccccc}
\hline                       
Star name                      & [C/Fe]  & [Mg/Fe]  & [Ca/Fe]   & [Cr/Fe]  & [Mn/Fe]  & [Ni/Fe]   & Ref \\
\hline
LAMOSTJ151003.74+305407.3      & 1.74    & $-$0.01  & $-$0.25   & $-$0.48  & $-$0.43  & $-$0.04   & 1 \\
                               & 0.81    & 0.21     & 0.36      & 0.11     & $-$0.15  & 0.07      & 2  \\        
\hline   
\end{tabular}}
 
References: 1. Our work, 2. Hayes et al. 2018 \\

\end{table*}
}

\subsection{Heavy element abundance analysis}
 \subsubsection{\textbf{The light s-process elements: Rb, Sr, Y, Zr}}
The spectral synthesis calculation of resonance line of Rb I at 7800.259 \r{A}
is used to derive the Rb abundance in both the stars. The Rb I 7947.597 \r{A} line was not
usable for the abundance estimation. The Rb hyperfine components are taken from Lambert \& Luck (1976).
It is mildly enhanced in LAMOSTJ091608.81+230734.6 with [Rb/Fe]$\sim$0.47,
while it is near-solar in LAMOSTJ151003.74+305407.3.

We could derive the strontium abundance only in LAMOSTJ091608.81+230734.6 
as no usable lines could be measured on the spectrum of the other star.
The Sr I 4607.327 {\rm \AA} line in LAMOSTJ091608.81+230734.6 returned 
a value [Sr/Fe]$\sim$0.97 from the spectral synthesis calculation.

\begin{figure}
\centering
\includegraphics[width=\columnwidth]{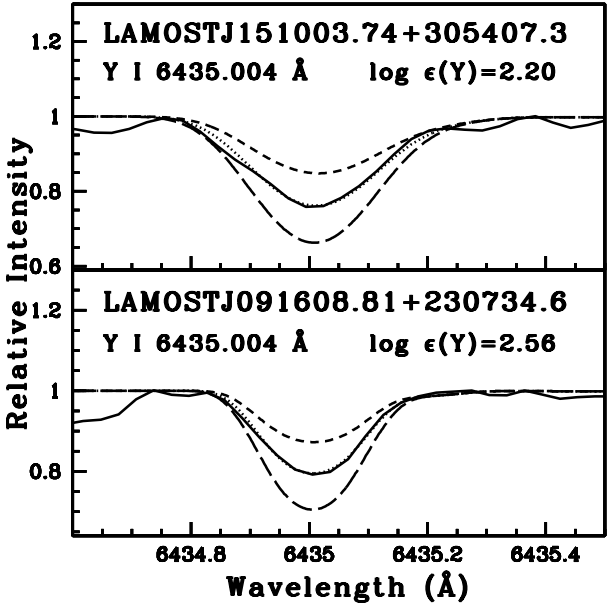}
\caption{ Synthesis of Y I line at 6435.004 {\rm \AA}. 
Dotted and solid lines represent synthesized and observed spectra respectively.
Short dashed and long dashed lines represent the synthetic spectra 
for $\Delta$[Y/Fe] = $-$0.3 and +0.3 respectively.} \label{Y6435}
\end{figure}

The abundance of yttrium is derived using the spectral synthesis calculation of Y I line at
6435.004 {\rm \AA} and equivalent width measurement of a few Y II lines in both 
the stars. Both the species give a value [Y/Fe]$\geq$1. 
The spectrum synthesis fits for Y I of the program stars are shown 
in Figure \ref{Y6435}. 

\begin{figure}
\centering
\includegraphics[width=\columnwidth]{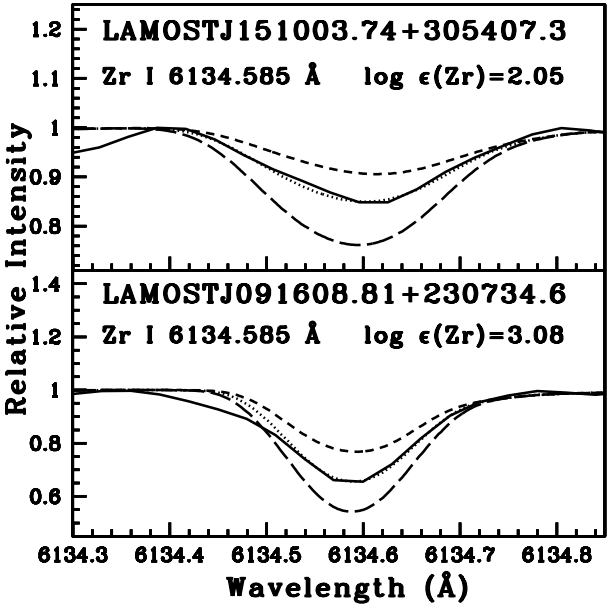}
\caption{ Synthesis of Zr I line at 6134.585 {\rm \AA}. 
Dotted and solid lines represent synthesized and observed spectra respectively.
Short dashed and long dashed lines represent the synthetic spectra 
for $\Delta$[Zr/Fe] = $-$0.3 and +0.3 respectively.} \label{Zr6134}
\end{figure}

The spectral synthesis calculation of Zr I line at 6134.585 {\rm \AA} and Zr II line
at 5112.297 {\rm \AA} are used to derive the zirconium abundance in the program 
stars. In both the cases, Zr is found to be enhanced with [Zr/Fe]$>$1. 
The spectrum synthesis fits for Zr I of the program stars are shown 
in Figure \ref{Zr6134}.

In the case of LAMOSTJ091608.81+230734.6, neutral lines of Y and Zr give a higher abundance than 
the singly ionized lines.

\subsubsection{\textbf{The heavy s-process elements: Ba, La, Ce, Pr, Nd}}
The spectral synthesis calculation of Ba II line at 5853.668 {\rm \AA} 
in LAMOSTJ091608.81+230734.6 and Ba II 6141.713 {\rm \AA} line  in 
 LAMOSTJ151003.74+305407.3 are used to derive the barium abundances.
La abundance is derived from the spectral synthesis calculation of 
 La II line at 5259.380 {\rm \AA} in both the stars. 
We could not detect any other useful lines due to lanthanum in the program stars for 
 the abundance determination.
 The abundances of Ce, Pr and Nd are derived using the measured equivalent 
 widths of several spectral lines due to singly ionized species of the respective elements.
 All these elements are found to be enhanced in both the stars with 
 [X/Fe]$>$1. 
 
\par Finally, we have estimated the [ls/Fe], [hs/Fe] and [hs/ls] ratios 
in the program stars. Here, ls and hs are light (Sr, Y and Zr) and heavy (Ba, La, Ce and Nd) 
s-process elements respectively. Also, we have estimated the mean abundance ratio of
s-process elements (Sr, Y, Zr, Ba, La, Ce, Nd ), [s/Fe], to find the 
s-process content.
 
\subsubsection{\textbf{The r-process elements: Sm, Eu} }
The Sm abundance is derived from the equivalent width measurement of 
Sm II lines listed in Table A2.
Both the stars show enhancement of Sm with [Sm/Fe]$\sim$1.34 and 
1.64 in LAMOSTJ091608.81+230734.6 and LAMOSTJ151003.74+305407.3 respectively.

The abundance of europium is estimated from the spectral synthesis 
calculation of Eu II 6645.064 {\rm \AA}. In LAMOSTJ091608.81+230734.6
Eu is enhanced with [Eu/Fe]$\sim$0.70 while the other star shows a value 1.14.

\section{Kinematic  Analysis}
The spatial velocity of a star in the solar neighborhood is measured with respect to
the Local Standard of Rest (LSR). The components of the spatial velocity are
$U_{LSR}$, $V_{LSR}$ and $W_{LSR}$; measured along the axes pointing 
towards the Galactic center, the direction of Galactic rotation and the 
North Galactic Pole respectively (Johnson \& Soderblom 1987).
Space velocity of the programme stars are calculated following the 
procedures in Bensby et al. (2003).
Components of the spatial velocity of the star with respect to LSR: \\
 \begin{center}
  $(U, V, W)_{LSR} =(U,V,W)+(U, V, W)_{\odot}$ km/s.
  \end{center}
where, $(U, V, W)_{\odot}$ is the solar motion with respect to LSR and its value is 
 (11.1, 12.2, 7.3) km/s (Sch{\"o}nrich et al., 2010) and  
   \begin{center}
  $\left[ \begin{array}{c}  U \\  V \\  W \end{array} \right] = B.\left[ \begin{array}{c} V_{r}  
  \\ k.\mu_{\alpha}/\pi  \\ k.\mu_{\delta}/\pi \end{array} \right]$ 
  \end{center}
  \noindent where, $B=T.A$, T is the transformation matrix connecting the Galactic coordinate 
  system and equatorial coordinate system and A is a coordinate matrix defined below \\
  \[T=\left[ \begin{array}{ccc} -0.06699 & -0.87276 & -0.48354 \\
   +0.49273 & -0.45035 & +0.74458 \\ 
   -0.86760 & -0.18837 & 0.46020\end{array} \right]\] 
  \[A=\left[ \begin{array}{ccc} Cos\alpha.Cos\delta & -Sin\alpha & -Cos\alpha.Sin\delta \\ 
  Sin\alpha.Cos\delta & Cos\alpha & -Sin\alpha.Sin\delta \\ 
  Sin\delta & 0 & Cos\delta\end{array} \right]\] 

  \noindent where $\alpha$ is the RA, $\delta$ the DEC, $V_{r}$ the radial velocity in km/s, 
  $k=4.74057 km/s$ equivalent of 1 AU in one year, $\mu_{\alpha}$ and $\mu_{\delta}$ 
  respectively the proper motions in RA and DEC in arcsec/year and $\pi$ the parallax in arcsec. 
The proper motion and parallax values are taken from GAIA DR2 (Gaia collaboration et al. 2018, https://gea.esac.esa.int/archive/) 
and SIMBAD astronomical database.
The spectroscopic velocity estimates have been used in this calculation. 
  
\noindent The total spatial velocity of the star: \\
 $V_{spa}^{2}=U_{LSR}^{2}+V_{LSR}^{2}+W_{LSR}^{2}$

\noindent Errors in the respective velocity components are calculated as follows:\\
 
 \[\left[ \begin{array}{c}  \sigma_{U}^{2} \\
  \sigma_{V}^{2}   \\
  \sigma_{W}^{2}  \end{array} \right]= 
  C.\left[ \begin{array}{c}  \sigma_{V_{r}}^{2} \\  
  (k/\pi)^{2}[\sigma_{\mu_{\alpha}}^{2}+(\mu_{\alpha}\sigma_{\pi}/\pi)^{2}] \\   
  (k/\pi)^{2}[\sigma_{\mu_{\delta}}^{2}+(\mu_{\delta}\sigma_{\pi}/\pi)^{2}] \end{array} \right] + \] 
   \[\frac{2\mu_{\alpha}\mu_{\delta}k^{2}\sigma_{\pi}^{2}}{\pi^{4}}.
  \left[ \begin{array}{c}  b_{12}.b_{13} \\ b_{22}.b_{23} \\
    b_{32}.b_{33} \end{array} \right]\]
 
\noindent where  $C_{ij}=b_{ij}^{2}$ and $\sigma'_{s}$ the errors in respective quantities. 
 
 \par The probability that a star belongs to the Galactic thin/thick disc or halo 
 population is calculated following  the procedure described in Mishenina et al. (2004), 
 Bensby et al. (2003, 2004) and Reddy et al. (2006), with assumption that the space velocities follow a 
 Gaussian distribution. \\

   $\begin{array}{c}
    P_{thin}=\frac{f_{1}.p_{1}}{P}, P_{thick}=\frac{f_{2}.p_{2}}{P}, 
    P_{halo}=\frac{f_{3}.p_{3}}{P} \\
   \end{array}$ \\
   
   $P=\Sigma f_{i}.p_{i}$ \\

  $ p_{i}= K_{i}.exp[\frac{-U_{LSR}^{2}}{2.\sigma_{U_{i}}^{2}}-\frac{(V_{LSR}-V_{ad})^{2}}
  {2.\sigma_{V_{i}}^{2}}-\frac{W_{LSR}^{2}}{2.\sigma_{W_{i}}^{2}}]$ \\

 $K_{i}= \frac{1}{(2.\pi)^{(3/2)}.\sigma_{U_{i}}.\sigma_{V_{i}}.\sigma_{W_{i}}}$; $i=1,2,3  $ \\

\noindent where $\sigma'_{s}$ are the dispersion in velocities, $V_{ad}$ the mean galactic 
rotation velocity for each stellar population relative to LSR and $f$ the fractional population. 
These values are taken from Reddy et al. (2006).The estimates of the total
spatial velocity and components of spatial velocity along with the probability estimates are given in 
Table \ref{kinematic analysis}. Our analysis shows that the star 
LAMOSTJ091608.81+230734.6 belongs to the thin disc population
and LAMOSTJ151003.74+305407.3 belongs to the Galactic halo population. 

{\footnotesize
\begin{table*}
\caption{Spatial velocity and probability estimates for 
the program stars} \label{kinematic analysis} 
\begin{tabular}{lccccccc} 
\hline                       
Star name                   & U$_{LSR}$             & V$_{LSR}$          & W$_{LSR}$ & V$_{spa}$  & p$_{thin}$ & p$_{thick}$ & p$_{halo}$ \\
                            & (kms$^{-1}$)          & (kms$^{-1}$)       & (kms$^{-1}$)  & (kms$^{-1}$) &          &             &       \\
\hline
LAMOSTJ091608.81+230734.6  & $-$12.14$\pm$2.99  & 1.57$\pm$1.41  & 2.71$\pm$2.95 & 12.53$\pm$1.98 & 0.99 & 0.01   & 0.00  \\
LAMOSTJ151003.74+305407.3  & $-$16.38$\pm$2.84  & $-$207.80$\pm$20.74 & $-$49.68$\pm$8.67  & 214.25$\pm$22.33 & 0.00  & 0.29  & 0.71 \\
\hline
\end{tabular} 
\end{table*}
}

\section{Binary status of the program stars}
For a precise identification of the source of carbon and heavy-element
enrichment in peculiar stars, it is very crucial to know their binary status.
A number of investigations dedicated to identify the binarity of the peculiar stars have been 
carried out to date. The precise radial velocity monitoring studies
have shown that most of the Ba and CH stars (McClure et al. 1980, McClure 1983, 1984, 
McClure \& Woodsworth 1990, Udry et al. 1998a,b, Lucatello et al. 2005, Jorissen et al. 2016) 
and a high fraction of CEMP-s/(r/s) stars(Lucatello et al. 2005, Starkenburg et al. 2014, 
Jorissen et al. 2016) are in binary systems. The compilation of Duquennoy \& Mayor
(1991) and Starkenburg et al. (2014)
have shown  the binary fraction of CEMP-s/(r/s) stars to be 100\%. 
However, a few  recent studies have reported a binary frequency of 
82$\pm$10\%  for CEMP-s/(r/s) stars (Hansen et al. 2016c) and 17$\pm$9\% for 
CEMP-no stars (Hansen et al. 2016b). 

\begin{figure}
\centering
\includegraphics[width=\columnwidth]{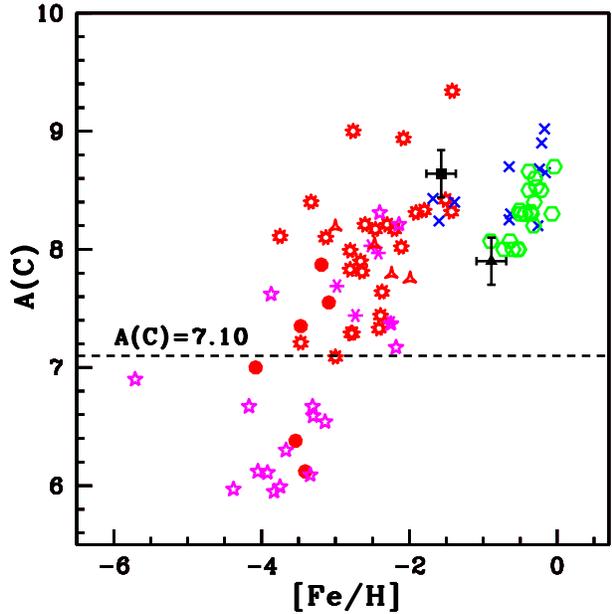}
\caption{Distribution of A(C) as a function of [Fe/H] for known (and likely)
binary and single stars.     
Red nine-sided stars, magenta six-sided crosses, red starred triangles, red filled circles
and magenta five-sided stars represent binary CEMP-s, single CEMP-s, binary CEMP-r/s, 
binary CEMP-no and single CEMP-no stars respectively from literature (Yoon et al. 2016).
All the red symbols corresponds to the binary CEMP stars and magenta symbols to single CEMP stars.
Blue crosses represent the binary CH stars from literature (Purandardas et al. 2019, 
Karinkuzhi \& Goswami 2014, 2015, Luck 2017). Binary Ba stars from literature 
(Shejeelammal et al. 2020, Karinkuzhi et al. 2018) are represented by green open hexagons. 
LAMOSTJ091608.81+230734.6 (filled triangle)
and LAMOSTJ151003.74+305407.3 (filled square). The dashed line indicate
the fiducial that separates binary and single stars.} \label{binarity}
\end{figure}

All these conclusions are based on the data available to each study. 
Information available in literature on the binary status of many CEMP stars is still very limited. 
Yoon et al.(2016) (and references therein) have done a compilation of the literature data for 305 CEMP stars
that includes 147 CEMP-s/(r/s) stars and 127 CEMP-no stars. Out of these, 
35 CEMP-s/rs stars and 22 CEMP-no stars have known binary status.
Earlier to this, Spite et al. (2013) observed for the first time the bimodality in the 
absolute carbon abundances among the CEMP stars, with CEMP-s stars populating the high-carbon 
region (A(C)$\sim$8.25) and CEMP-no stars populating the low-carbon region (A(C)$\sim$6.5). 
Later, this bimodality is confirmed by Bonifacio et al (2015) and Hansen et al (2015a) 
for an extended sample of CEMP stars.
Yoon et al. (2016) investigated anew the distribution of absolute carbon abundance, A(C), in terms of the 
metallicity for these CEMP stars. Their analysis have shown that the fiducial line at 
A(C)=7.1 in the [Fe/H] versus A(C) diagram could well separate the binary nature of 
all the CEMP-s/(r/s) stars and majority of CEMP-no stars.  
Despite of a few outliers, majority of the binary stars lie above this line. 
Thus this diagram can be used as a tool to derive clues on the binary status of
the CEMP stars as well as the origin of their chemical peculiarities. 
We have used such a figure to understand the binary nature of our program stars. 
The distribution of absolute carbon abundance with metallicity
for different classes of chemically peculiar stars are shown in 
figure \ref{binarity}. In this figure, both the program stars lie in the 
region occupied by binary stars, indicating that our program stars are likely binaries.
In addition, the variation in radial velocity estimate for LAMOSTJ151003.74+305407.3
by $\sim$4 kms$^{-1}$ also suggests that this object could be a binary star.

\section{Discussion}
\begin{figure}
\centering
\includegraphics[width=0.8\columnwidth, height= 0.8\columnwidth]{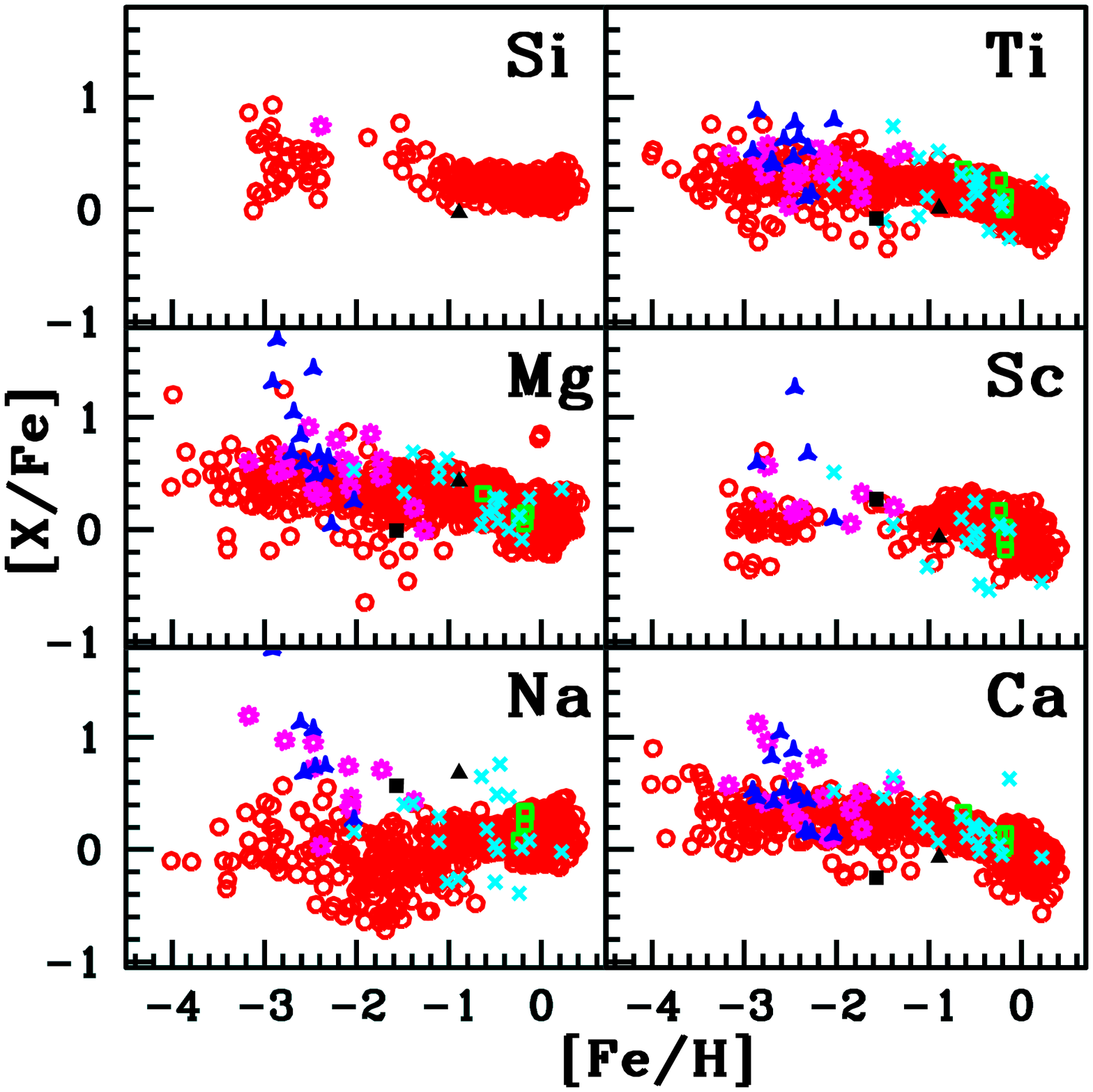}
\includegraphics[width=0.8\columnwidth, height= 0.8\columnwidth]{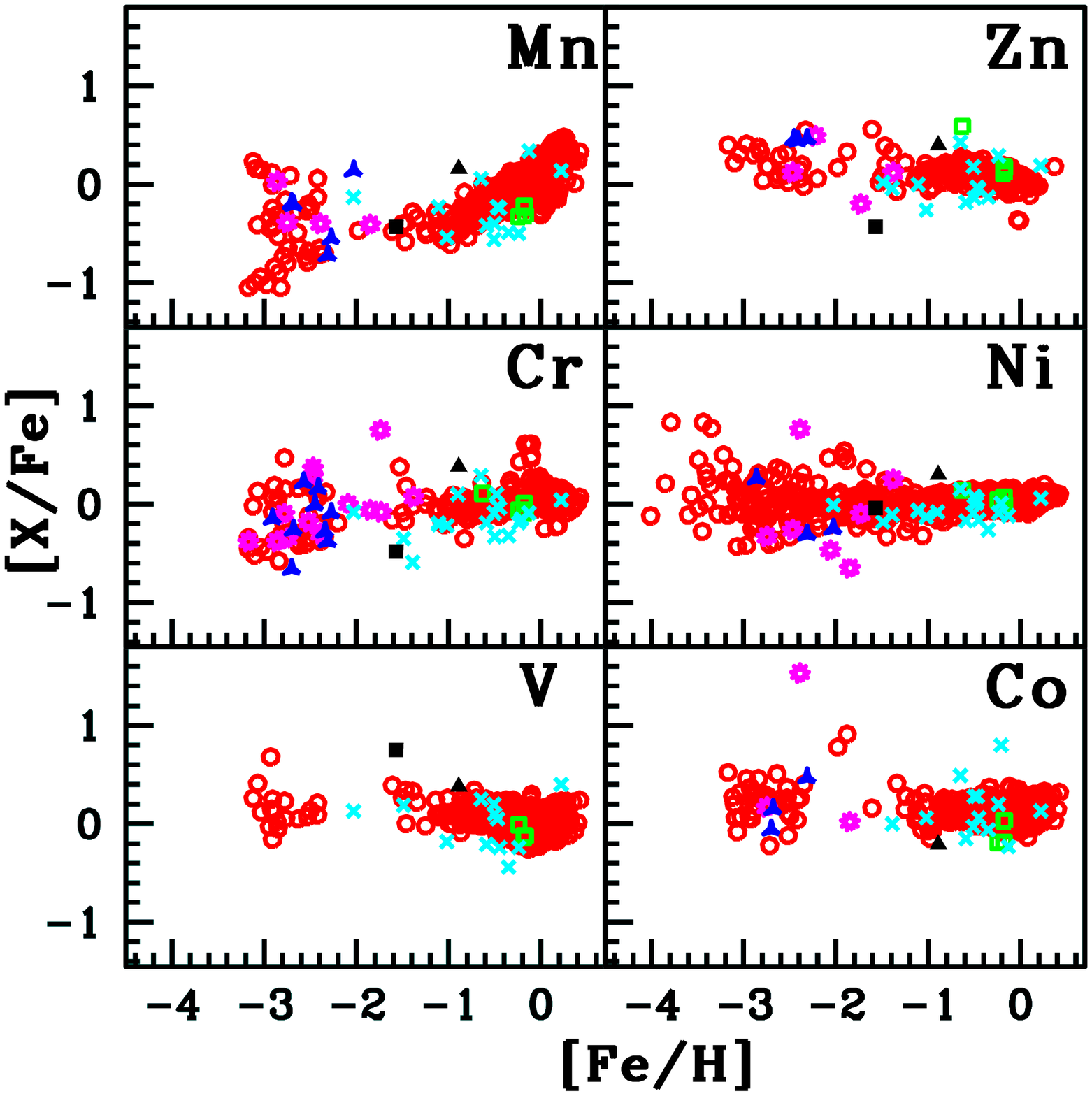}
\caption{Observed [X/Fe] ratios of the light elements in the  
program stars with respect to metallicity [Fe/H].   
Red open circles correspond to normal giants from literature (Honda et al. 2004, Venn et al. 2004, 
Aoki et al. 2005, 2007, Reddy et al. 2006, Luck \& Heiter 2007, Hansen et al. 2016a, Yoon et al. 2016).
Magenta nine-sided stars and blue starred triangles represent CEMP-s and
CEMP-r/s stars respectively from literature (Masseron et al. 2010). Cyan crosses and 
green open squares represent giant and sub-giant CH stars respectively from literature 
(Vanture 1992, Karinkuzhi \& Goswami 2014, 2015, Goswami et al. 2016).  LAMOSTJ091608.81+230734.6 (filled triangle)
and LAMOSTJ151003.74+305407.3 (filled square).} \label{light_elements}
\end{figure}

\begin{figure}
\centering
\includegraphics[width=0.8\columnwidth, height= 0.8\columnwidth]{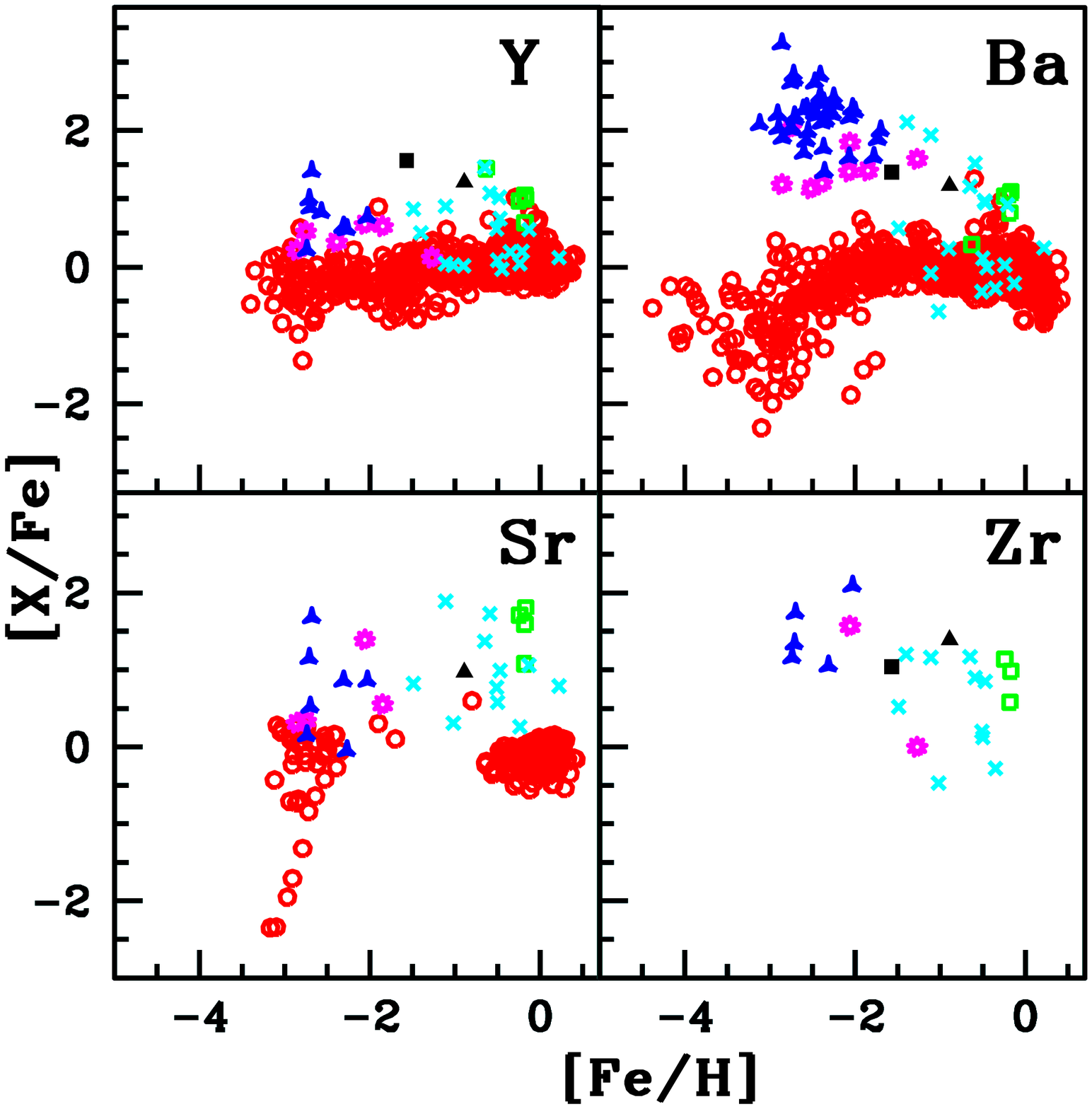}
\includegraphics[width=0.8\columnwidth, height= 0.8\columnwidth]{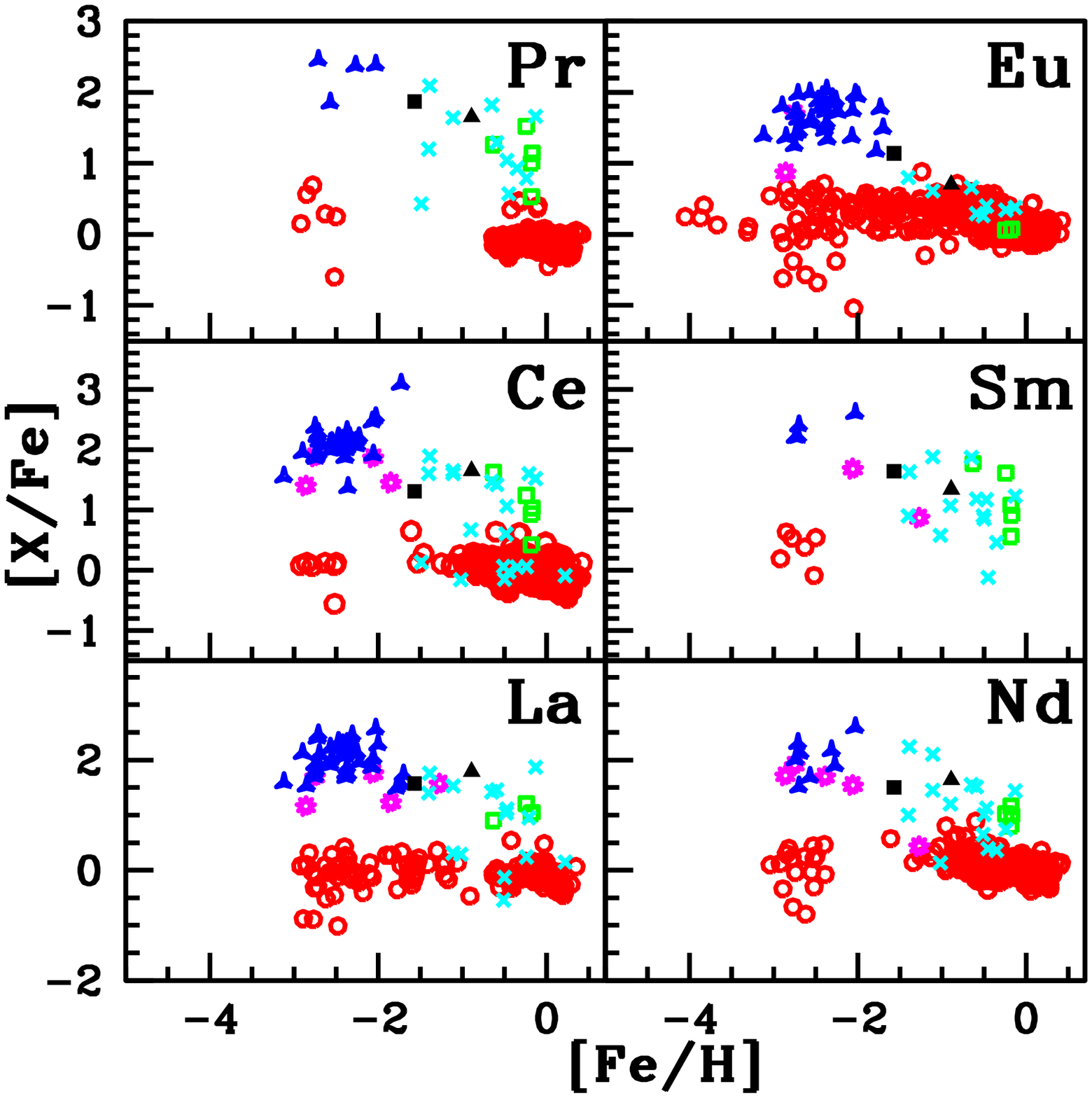}
\caption{Observed [X/Fe] ratios of the heavy elements in the 
program stars with respect to metallicity [Fe/H].   
Symbols have same meaning as in Figure \ref{light_elements}.} \label{heavy_elements}
\end{figure}

From the multiple line indices measured from the stellar spectra 
from LAMOST DR2, Ji et al. (2016) have identified the objects 
LAMOSTJ091608.81+230734.6 and LAMOSTJ151003.74+305407.3 to be CH stars. 
Our detailed abundance analysis confirms LAMOSTJ091608.81+230734.6 
to be a CH star and LAMOSTJ151003.74+305407.3 to be a CEMP -r/s star.
The estimated abundances of light elements from Na to Zn 
are quite similar to the normal giants following the 
Galactic trend as shown in Figure \ref{light_elements}.
The estimated abundances of heavy elements however show 
enhancement compared to their counterparts in other normal 
stars (Figure \ref{heavy_elements}).

\par
\begin{figure}
\centering
\includegraphics[width=\columnwidth]{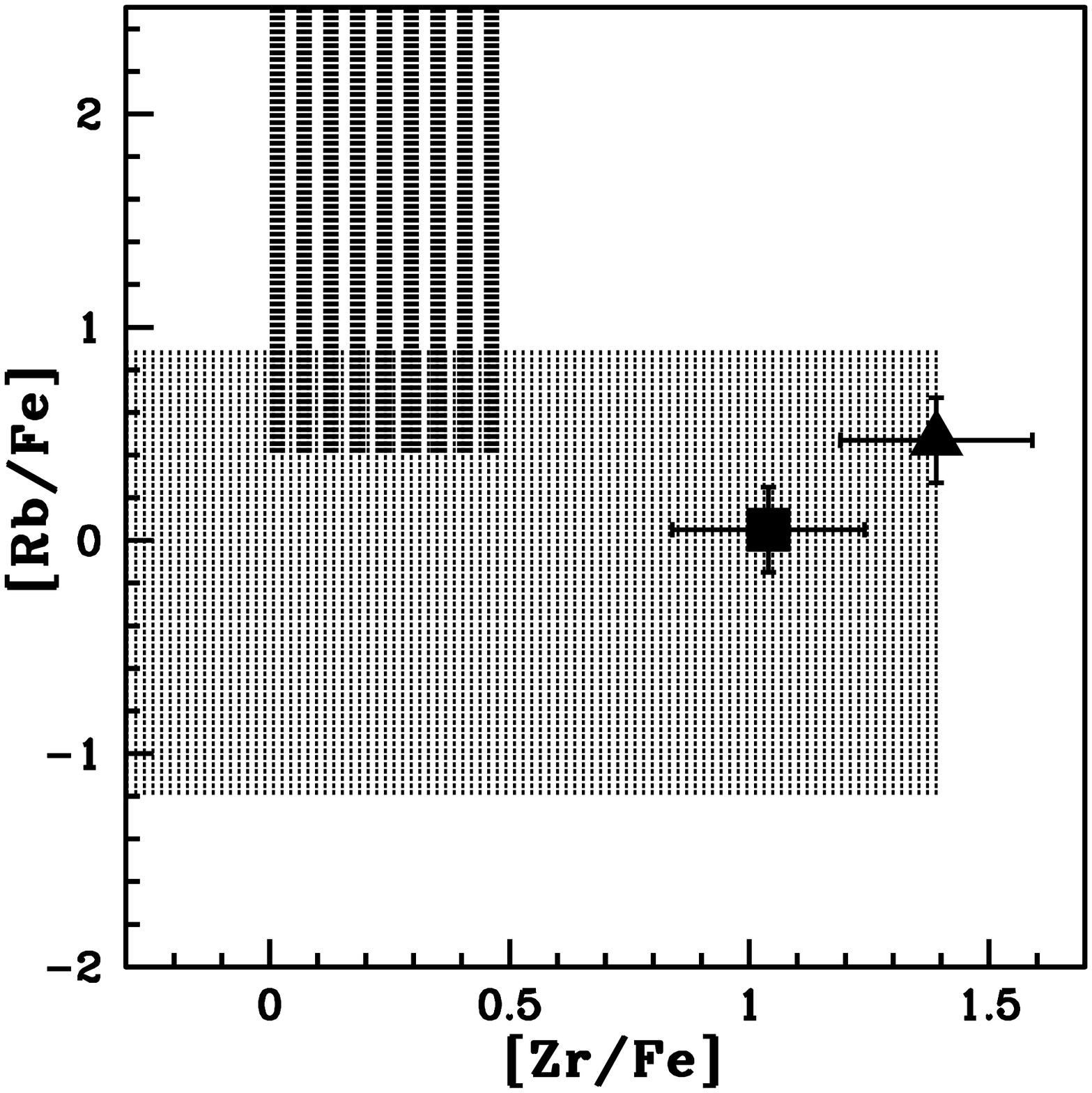}
\caption{The observed [Rb/Fe] and [Zr/Fe].
LAMOSTJ091608.81+230734.6 (filled triangle)
and LAMOSTJ151003.74+305407.3 (filled square). 
The shaded regions represent the observed ranges of 
Zr and Rb in intermediate-mass (short-dashed lines) 
and low-mass (dots) AGB stars of the 
Galaxy and the Magellanic Clouds (van Raai et al. 2012). 
The Rb and Zr abundances in the program stars are consistent 
with that of low-mass AGB stars.} \label{rbzr}
\end{figure}

 The neutron density dependent [Rb/Zr] ratio has been estimated 
in order to see the nature of companion star. For this ratio, the 
AGB models predict a negative value in low-mass AGB stars 
(M$\leq$ 3 M$_{\odot}$) and a positive value  in intermediate-mass 
AGB stars (M$\geq$ 4 M$_{\odot}$) (Abia et al. 2001, 
van Raai et al. 2012, Karakas et al. 2012).
This trend has been observed among the AGB stars in the Galaxy and 
Magellanic Clouds (Plez et al. 1993, Lambert et al. 1995, 
Abia et al. 2001,  Garc\'ia-Hern\'andez et al. 2006, 2007, 2009).
The estimated values of this ratio (Table \ref{hs_ls}) are shown in 
Figure \ref{rbzr}. The range of Rb and Zr observed in the
Galactic and Magellanic Cloud low- and intermediate-mass AGB stars (shaded regions) 
are also shown in the same figure.
\\

{\footnotesize
\begin{table*}
\caption{Estimates of  [ls/Fe], [hs/Fe], [s/Fe], [hs/ls], [Rb/Zr], [Ba/Eu] and  C/O} \label{hs_ls}
\begin{tabular}{lcccccccc}
\hline                       
Star name                      & [Fe/H]   & [ls/Fe] & [hs/Fe]  & [s/Fe] & [hs/ls] & [Rb/Zr]  & [Ba/Eu]    & C/O\\ 
\hline
LAMOSTJ091608.81+230734.6     & $-$0.89   & 1.20    & 1.57     & 1.41   & 0.37    & $-$0.92  & 0.49       & 1.38   \\  
LAMOSTJ151003.74+305407.3     & $-$1.57   & 1.30    & 1.44     & 1.67   & 0.14    & $-$0.99  & 0.25       & -- \\     
\hline
\end{tabular}
\end{table*}
}

\noindent {\small\textbf{LAMOSTJ091608.81+230734.6:}}  
The object is found to be metal-poor with a metallicity of 
$-$0.89. Eventhough we could not find its position on HR diagram,
its estimated log g value suggests the star to be on RGB (Allen \& Barbuy 2006).
The star shows a mild over abundance of carbon 
with [C/Fe]$\sim$0.36. Similar values of carbon in CH stars
have been reported in literature (Purandardas et al. 2019, Goswami et al. 2016,
Karinkuzhi \& Goswami 2015, Vanture 1992). Nitrogen enhancement compared 
to carbon in the star, along with the low $^{12}$C/$^{13}$C ratio 
of $\sim$ 8.67, suggest the CN processing and the First Dredge-Up (FDU)
during the giant phase. Further, with its estimated values 
of mean s-process abundance ratio,
[s/Fe]$\sim$1.41, and C/O$\sim$1.38, this star can be included in the 
CH giant category. The star shows negative value for the [Rb/Zr], which indicates 
a low-mass companion AGB as expected for the CH stars (Figure \ref{rbzr}). The estimated [hs/ls] ratio 
is $\sim$0.37, indicating the over abundance of second-peak s-process elements over the 
first peak, as normally seen in most of the CH stars. 
Our kinematic analysis shows that this star belongs to Galactic thin disc 
population.

\begin{figure}
\centering
\includegraphics[width=\columnwidth]{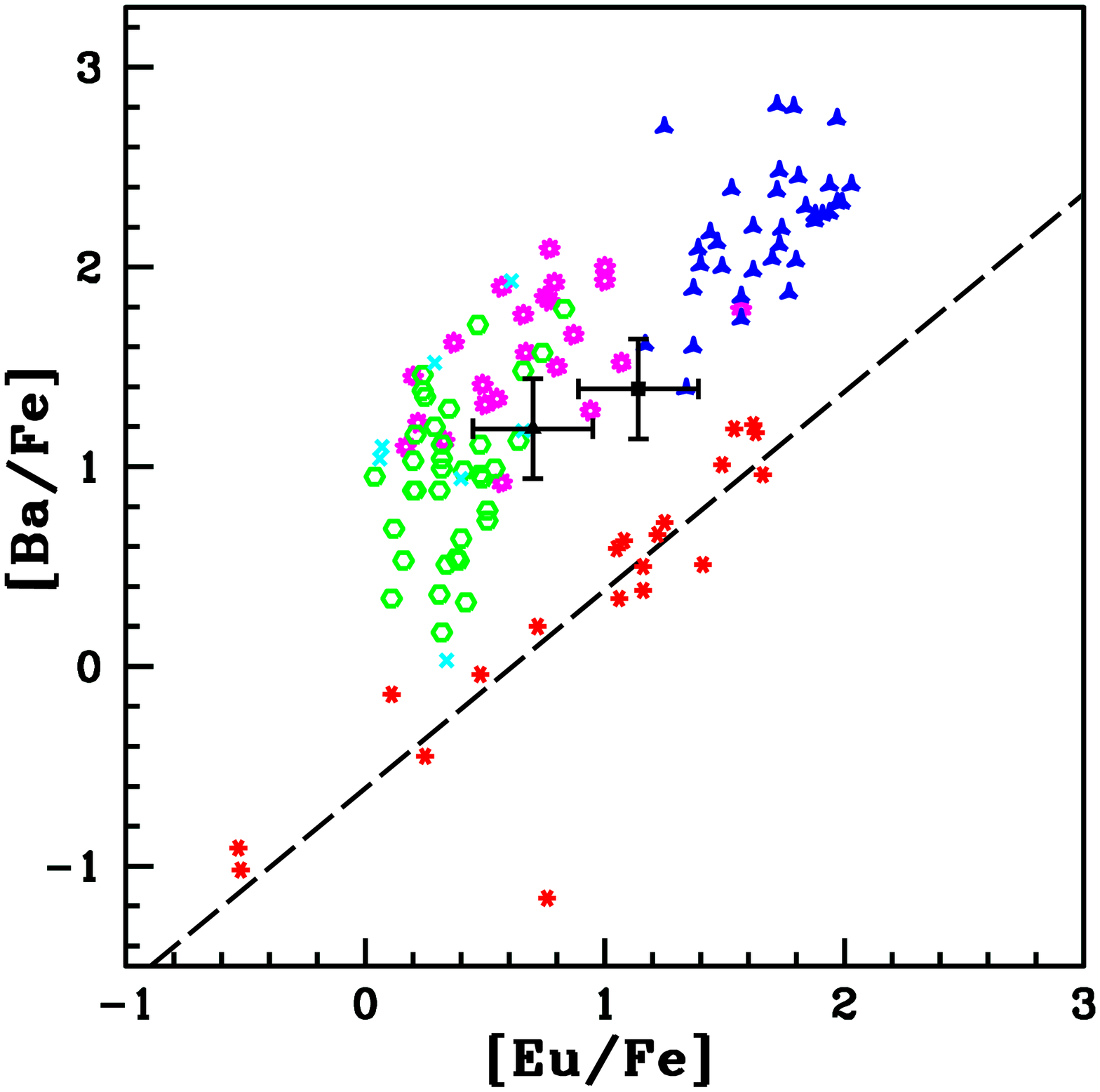}
\caption{Observed [Eu/Fe] and [Ba/Fe] ratios for 
different classes of chemically peculiar stars.   
Magenta nine-sided stars, blue starred triangles and red six-sided crosses represent CEMP-s
CEMP-r/s and r (including both CEMP-r and rI/rII stars) stars respectively 
from literature (Masseron et al. 2010). Cyan crosses represent 
CH stars from literature (Vanture 1992, Karinkuzhi \& Goswami 2014, 2015, Goswami et al. 2016). 
Green open hexagons represent the Ba stars from literature
(Shejeelammal et al. 2020, Yang et al. 2016, Allen \& Barbuy 2006). LAMOSTJ091608.81+230734.6 (filled triangle)
and LAMOSTJ151003.74+305407.3 (filled square). The dashed line is the least-square 
fit to the observed abundances in r-stars.} \label{Ba_Eu}
\end{figure}

Figure \ref{Ba_Eu} compares the observed abundance ratios [Ba/Fe] and [Eu/Fe],
the representative s- and r- process elements respectively, in different classes 
of chemically peculiar stars. The star LAMOSTJ091608.81+230734.6 falls in the region 
occupied by  other CH stars, Ba stars and CEMP-s stars. These three
classes of stars are known to have the same origin of s-process enhancement, 
pollution from a binary low-mass AGB companion. \\    

\begin{figure}
\centering
\includegraphics[width=\columnwidth]{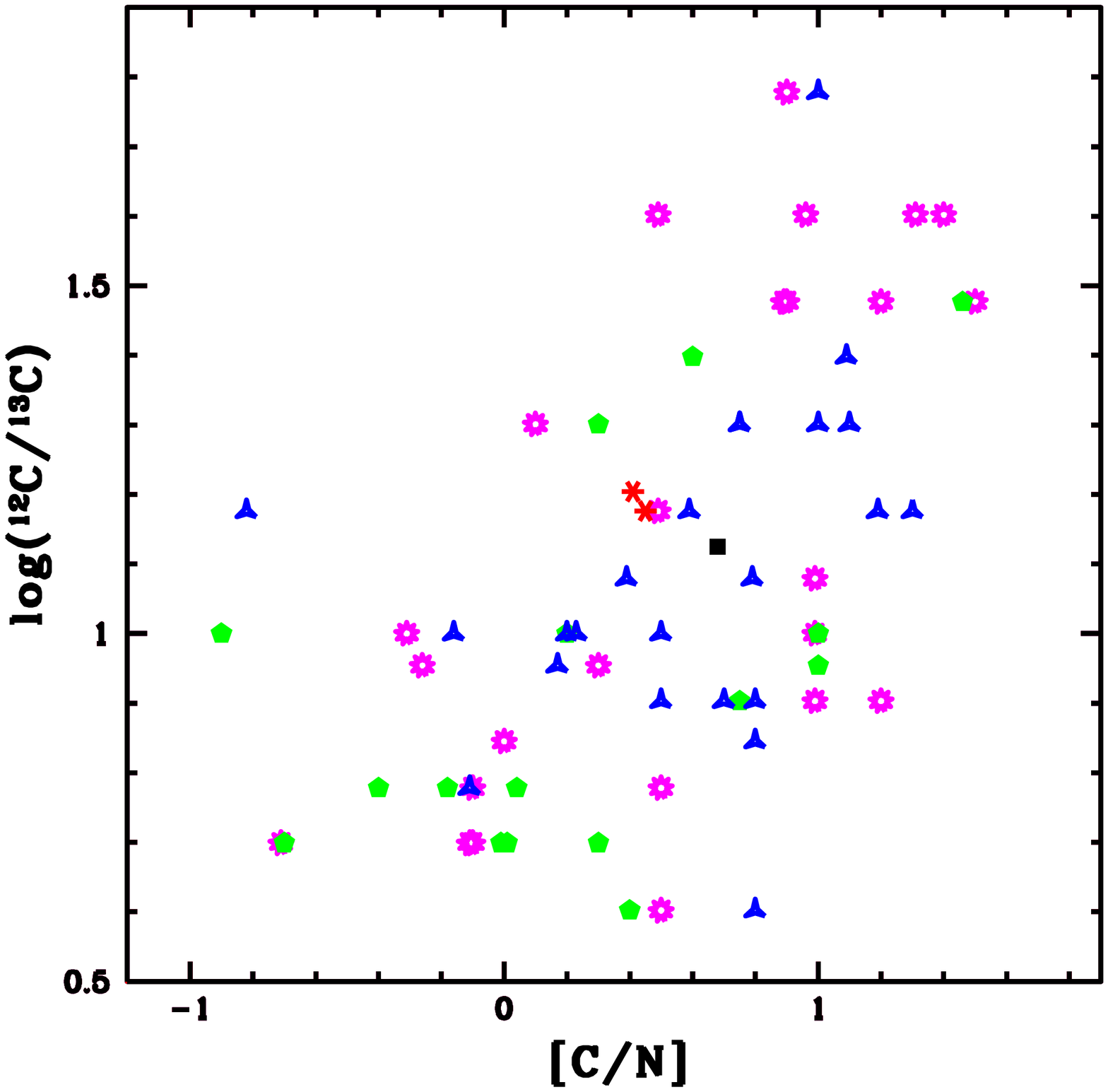}
\caption{Observed [C/N] and log($^{12}$C/$^{13}$C) ratios 
of the CEMP stars. 
Magenta nine-sided stars, red six-sided crosses, blue starred triangles 
and green filled pentagons represent CEMP-s, CEMP-r, CEMP-r/s and
CEMP-no stars respectively from literature (Masseron et al. 2010). 
LAMOSTJ151003.74+305407.3 (filled square).} \label{12c13c_cn}
\end{figure}

\noindent {\small\textbf{LAMOSTJ151003.74+305407.3:}}  
This star is found to be metal-poor with a metallicity 
[Fe/H] ${\sim}$ $-$1.57
and enhanced in carbon with [C/Fe]$>$1. With its estimated 
value of [Ba/Fe]$\sim$1.39 and [Ba/Eu]$\sim$0.25, along with the
enhanced carbon abundance, the star satisfies to be a CEMP-r/s star (Beers \& Christlieb, 2005, Abate et al. 2016).
From the position of the star on the HR diagram, it is found to be a giant.
The abundances of heavy elements in this star are well within the range observed  
in  other CEMP-r/s stars (Figure \ref{heavy_elements}). In 
[C/N] versus log($^{12}$C/$^{13}$C) space, this star occupies the same region 
as the CEMP stars (Figure \ref{12c13c_cn}). 

\par
\begin{figure}
\centering
\includegraphics[width=\columnwidth]{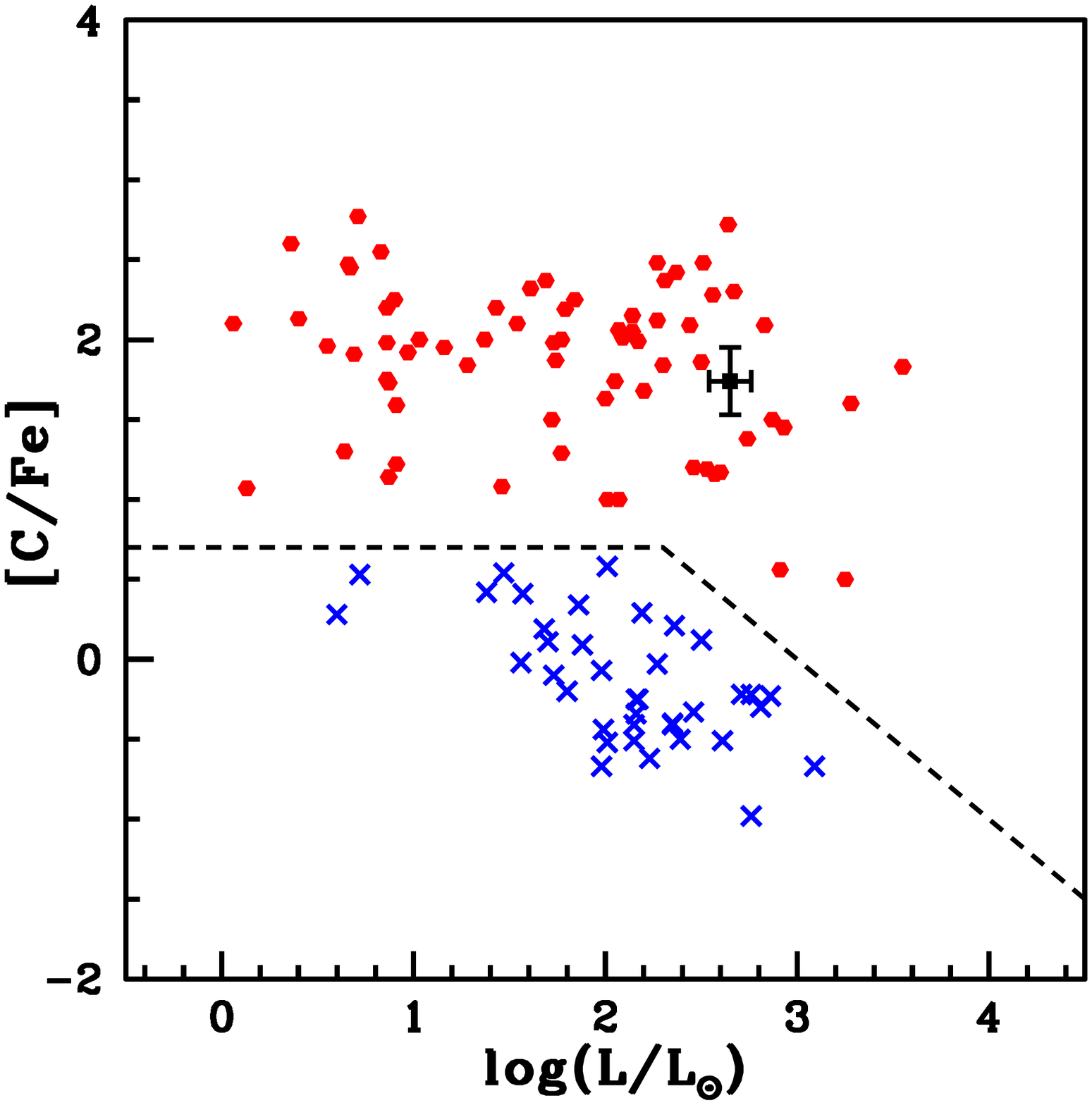}
\caption{Observed [C/Fe] ratios as a function of luminosity estimated from the effective 
temperature.   
Red filled hexagons represent CEMP stars from literature 
(Aoki et al. 2007 and references therein, 
Purandardas et al. 2019, Goswami et al. 2016).
Blue crosses represent carbon-normal metal-poor stars from literature 
(Aoki et al. 2005, 2007, Cayrel et al. 2004, Honda et al. 2004). 
LAMOSTJ151003.74+305407.3 (filled square). 
The dashed line indicates the dividing
line between CEMP and the carbon-normal metal-poor
stars.} \label{carbon_luminosity}
\end{figure}

 From the study conducted by Aoki et al.(2007) for 22 metal-poor stars 
($-$3.3$\leq$[Fe/H]$\leq$-1.0) exhibiting strong 
CH and/or C$_{2}$ molecular band, they came up with a new empirical definition for CEMP stars 
in terms of [C/Fe] and log(L/L$_{\odot})$. Based on this definition, there exist a clear division between 
the CEMP stars and the carbon-normal metal-poor stars in the
log(L/L$_{\odot}$) versus [C/Fe] plane. We have demonstrated this in Figure \ref{carbon_luminosity}. 
The star LAMOSTJ151003.74+305407.3 is also shown in the same figure. This star 
falls in the region occupied by the CEMP stars. 

The Figure \ref{Ba_Eu} shows the position of the star LAMOSTJ151003.74+305407.3
in the [Eu/Fe] - [Ba/Fe] space. The star is not placed in the bulk of the CEMP-r/s stars,
however, there are a few CEMP-r/s stars in the same region as LAMOSTJ151003.74+305407.3.

\par With the knowledge that r-process and s-process are ascribed to different astrophysical 
sites (Burbidge et al. 1957), several formation scenarios for CEMP-r/s stars 
have been proposed in literature (Cohen et al. 2003, Qian \& Wasserburg 2003, 
Zijlstra 2004, Barbuy et al. 2005, Wanajo et al. 2006,
Jonsell et al. 2006, Bisterzo et al. 2011 and references therein), 
most of them suggesting different independent processes for the r- and s- peculiarity. 
One scenario describes that the CEMP-r/s star could have been a secondary in a binary 
system formed out of r-process enriched ISM and its s-elements and carbon enhancement is 
due to the later pollution from the AGB companion through the mass-transfer 
mechanism (Hill et al. 2000, Cohen et al. 2003, Ivans et al. 2005). 
But this could not successfully explain the observed frequency of CEMP-r/s stars. 
The study by Barklem et al. (2005) have revealed that an  
order of 1\% of population II stars are CEMP-r/s stars.
Cohen et al (2003) have discussed another scenario that invokes a triple system in which
the star could have been a least massive tertiary, polluted first with the
r-elements from the massive primary exploded as supernova and later polluted
with s-elements by the secondary star evolved into an AGB . This hypothesis had been discarded,
as such a dynamically stable tertiary system is unlikely to exist. 
Accretion Induced Collapse (AIC) in a binary system have been 
suggested by Qian \& Wasserburg (2003) and Cohen et al. (2003), but discarded
since it is physically uncertain with the existing neutrino theories (Qian \& Woosley 1996).
Another scenario is a binary picture in which the primary star evolved through AGB
contributed the s-rich material to the secondary star, and later 
explodes as Type 1.5 supernova (Zijlstra 2004, Wanajo et al. 2005)
depositing r-material on the surface of the secondary. 
Abate et al. (2016) have calculated the frequency of CEMP-r/s stars among the 
CEMP-s stars for all these formation scenarios. The theoretical frequency 
predicted in most of the scenarios underestimate the observed 
frequency ($\sim$54\% in their sample) atleast by a factor of five.
The simulation based on the hypothesis of independent enrichment of s- and r- elements
could predict a frequency ($\sim$22\%) that approaches the observations,
however, it fails to reproduce the observed correlation of observed Ba and Eu 
abundances in CEMP-r/s stars.  
The simulations of Jonsell et al. (2006) for a high neutron density s-process
in AGB star in a binary system also could not reproduce the observed 
abundance pattern in CEMP-r/s stars.

\par Allen et al. (2012) have claimed from their analysis for a sample of 
CEMP stars that both CEMP-s and CEMP-r/s stars have same astrophysical origin.
A modified neutron-capture process called intermediate neutron-capture process (i-process)
in AGB stars, 
first proposed by Cowan \& Rose (1977), has been invoked recently 
(Dardelet et al. 2014, Hampel et al 2016, 2019, Hansen et al. 2016c) to explain 
the observed abundance trend of CEMP-r/s stars in the context of 
binary-mass transfer scenario. When a substantial amount of hydrogen rich-material
is mixed into the intershell region of the evolved red-giant stars (Proton Ingestion Episodes, PIE) 
undergoing helium shell flash, a significantly high neutron density of 
the order N$_{n}$ $\sim$ 10$^{15}$ - 10$^{17}$ cm$^{-3}$
(intermediate between s- and r- processes) can be produced (Cowan \& Rose 1977). 
A number of sites have been proposed for the PIEs in order for the i-process nuclesynthesis 
to take place. Recent simulations have shown that the neutron-densities of the order of 
10$^{12}$ - 10$^{15}$ cm$^{-3}$ can be achieved in very low-metallicity (z $\leq$ 10$^{-4}$),
low-mass (M $\leq$ 2M$_{\odot}$) AGB stars (Fujimoto et al. 2000, Campell \& Lattanzio 2008, 
Lau et al. 2009, Cristallo et al. 2009, Campbell et al. 2010, Stancliffe et al. 2011). 
A similar neutron density is achieved during 
the dual core flash (H flash following the PIEs during the He flash) in low-mass, 
extremely low-metallicity (z $\leq$ 10$^{-5}$) stars (Fujimoto et al. 1990, Hollowell et al. 1990, Lugaro et al. 2009); 
in the very-late thermal pulses of post-AGB stars (Herwig et al. 2011, 2014 
Bertolli et al. 2013, Woodward et al. 2015);
during the thermal pulses of low-metallicity super-AGB stars (Doherty et al. 2015, Jones et al. 2016);
in Rapidly Accreting White Dwarfs (RAWD) (Denissenkov et al. 2019). 
Dardelet et al. (2014), Hampel et al. (2016) and Hampel et al. (2019) considered the possibility of 
i-process in their simulation to see whether the CEMP-r/s phenomena could be explained 
on the basis of s- and r- process elements  
produced at a single stellar site. The i-process models could satisfactorily reproduce 
the abundance patterns of twenty CEMP-r/s stars (Hampel et al. 2016)
and seven low-Pb  Magellanic post-AGB stars (Hampel et al. 2019).

\par Hampel et al. (2016) used the single-zone nuclear network calculations 
to simulate the properties of the intershell region of low-mass (1 M$_{\odot}$),
low-metallicity (z = 10$^{-4}$) AGB star and studied the neutron-capture 
nucleosynthesis under the influence of different constant neutron densities
ranging from 10$^{7}$ to 10$^{15}$ cm$^{-3}$. The physical input conditions of intershell
region are adapted from Stancliffe et al. (2011) and the composition from Abate et al. (2015b).   
The considered temperature and density of the intershell region are 1.5 $\times$ 10$^{8}$ K and 
1600 g cm$^{-3}$ respectively. Compared to the classical s-process, at i-process neutron densities, 
this simulation resulted in an increased production of heavy s-process and r-process elements, while 
similar abundances of light s-process elements, as typically observed for CEMP-r/s stars 
(Abate et al. 2015a, Hollek et al. 2015).  
A range of temperatures (1.0 $\times$ 10$^{8}$ - 2.2 $\times$ 10$^{8}$ K) and densities 
(800 - 3200 g cm$^{-3}$) have been tested in this simulation, which did not produce significant 
changes in the result. 

\par We have compared the observed heavy elemental abundance ratios of 
the star LAMOSTJ151003.74+305407.3 with the model yields, [X/Fe], of
Hampel et al. (2016) for a range of neutron densities, n$\sim$10$^{9}$ - 10$^{15}$ cm$^{-3}$. 
A further dilution of the accreted material can occur on the surface of the CEMP-r/s 
stars just like similarly formed CEMP-s stars (Stancliffe et al. 2007, 2013).    
The neutron-density responsible for the observed abundance in this star
is derived by fitting the observed abundance with the dilution factor incorporated parametric model function: \\

\begin{figure}
\centering
\includegraphics[width=\columnwidth]{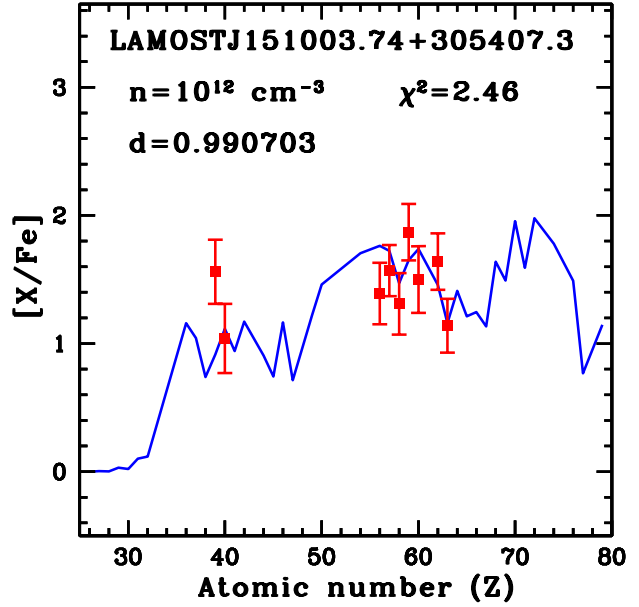}
\caption{The best fit obtained for the parametric model function is represented by  
the solid line. The squares with error bars are the observed abundances in 
LAMOSTJ151003.74+305407.3.} \label{parametric_j151003}
\end{figure}

\noindent X = X$_{i}$ . (1-d) + X$_{\odot}$ . d \\  

\noindent where X is the final abundance, X$_{i}$ is the i-process abundance, 
X$_{\odot}$ is the solar-scaled abundance and d is the dilution factor.
The best fit obtained along with the neutron density and dilution factor
is shown in Figure \ref{parametric_j151003}. The neutron density 
responsible for the observed abundance of this star is found to be 
n$\sim$10$^{12}$ cm$^{-3}$, if we assume a single stellar site for the production of
the observed neutron-capture abundance pattern. 

From the kinematic analysis, we find that  
this object belongs to the Galactic halo with a probability of 71\%. 
The spatial velocity estimate of the star is similar to the typical velocity 
of halo objects, V$_{spa}$$>$180 kms$^{-1}$ (Chen et al. 2004).
Also it satisfies the criteria, [Fe/H]$<$$-$0.90 and V$_{LSR}$$<$$-$120 km s$^{-1}$
(Eggen 1997) to be a halo object.

\section{CONCLUSIONS}
The results from a detailed high-resolution spectroscopic analysis of two 
carbon stars identified from LAMOST DR2 are presented. Both the
objects are identified to be CH stars by Ji et al. (2016).
Our analysis shows that the object LAMOSTJ151003.74+305407.3 is a CEMP-r/s 
star, while the object LAMOSTJ091608.81+230734.6 is a CH star. 

Although, a few light element abundances are available in literature for LAMOSTJ151003.74+305407.3, 
we have presented the first time detailed abundance analysis for both the stars. 
We have estimated the abundances of 26 elements along with the carbon isotopic ratio $^{12}$C/$^{13}$C. 

Our analysis based on the neutron-density dependent [Rb/Zr] ratio
confirms  low-mass for the former AGB companion of the program stars.
The kinematic analysis shows, LAMOSTJ091608.81+230734.6 belongs to 
the Galactic disc, and  LAMOSTJ151003.74+305407.3 belongs to Galactic 
halo population.

An i-process parametric model based analysis performed for the CEMP-r/s star  
LAMOSTJ151003.74+305407.3,  
yields a neutron density n$\sim$10$^{12}$ cm$^{-3}$ at the neutron-capture
nucleosynthesis site, may indicate that the i-process in the companion AGB star
be responsible for its observed abundance pattern.
 
 \section{ACKNOWLEDGMENT}
 We thank the staff at IAO and at the remote control station at 
CREST, Hosakotte for assisting during the observations.
Funding from the DST SERB project No. EMR/2016/005283 is gratefully 
acknowledged. We are thankful to the referee, Dr. Luca Sbordone, for useful 
comments and suggestions that have considerably improved the paper. We are thankful to Melanie Hampel for providing us with 
the i-process yields in the form of number fractions and Partha Pratim
Goswami for generating the model fits used in Figure 15.
This work made use of the SIMBAD astronomical database, operated
at CDS, Strasbourg, France, and the NASA ADS, USA.
This work has made use of data from the European Space Agency (ESA) 
mission Gaia (https://www.cosmos.esa.int/gaia), processed by the Gaia 
Data Processing and Analysis Consortium 
(DPAC, https://www.cosmos.esa.int/web/gaia/dpac/consortium).
Based on data collected using  HESP \\

\noindent
{\bf Data Availability}\\
The data underlying this article will be shared on reasonable request to the corresponding author.

{}

\section*{Appendix}
{\footnotesize
\begin{table*}
{\bf Table A1 : Equivalent widths (in m\r{A}) of Fe lines used for deriving 
atmospheric parameters}
\resizebox{\textwidth}{!}{
\begin{tabular}{lccccccc}
\hline                       
Wavelength(\r{A}) & El & $E_{low}$(eV) & log gf & LAMOSTJ091608.81+230734.6 & LAMOSTJ151003.74+305407.3 & Ref  \\ 
\hline 
4445.471	&	Fe I & 0.087	&	 -5.380	&	 77.50(6.44)	&	 -		&	 1  \\
4484.220	&	 &  3.603	&	 -0.720	&	 91.70(6.40)	&	 88.1(5.72)	&	 1  \\
4489.739	&	 &  	 0.121	&	 -3.966	&	 133.5(6.39	&	 -		&	 1  \\
4619.288	&	 &  	 3.603	&	 -1.120	&	 84.2(6.60)	&	 76.7(5.95)	&	 1  \\
4635.846	&	 &  	 2.845	&	 -2.420	&	 56.2(6.45)	&	 -		&	 1  \\
4637.503	&	 &  	 3.283	&	 -1.390	&	 86.3(6.53)	&	 -		&	 1  \\
4643.464	&	 &  	 3.654	&	 -1.290	&	 70.9(6.54)	&	 -		&	 1  \\
4690.138	&	 &  	 3.686	&	 -1.640	&	 53.7(6.60)	&	 -		&	 1  \\
4882.143	&	 &  	 3.417	&	 -1.640	&	 82.1(6.80)	&	 55.7(5.94)	&	 1  \\
4907.732	&	 &  	 3.430	&	 -1.840	&	 61.7(6.62)	&	 -		&	 1  \\
4908.031	&	 &  	 4.217	&	 -1.396	&	 39.8(6.58)	&	  -		&	 2  \\
4917.229	&	 &  	 4.191	&	 -1.180	&	 35.7(6.38)	&	 -		&	 1  \\
4924.770	&	 &  	 2.278	&	 -2.220	&	 111(6.64)	&	 112(5.85)	&	 1	  \\
4939.687	&	 &  	 0.859	&	 -3.340	&	 142(6.64)	&	 173(6.02)	&	 1  \\
4967.890	&	 &  	 4.191	&	 -0.622	&	 82.8(6.7)	&	 -		&	 2  \\
4969.917	&	 &  	 4.216	&	 -0.710	&	 56.7(6.31)	&	 -		&	 1  \\
4985.253	&	 &  	 3.930	&	 -0.560	&	 84.8(6.36)	&	 -		&	 2  \\
5022.236	&	 &  	 3.984	&	 -0.530	&	 80.8(6.31)	&	 96(6.05)	&	 1  \\
5028.126	&	 &  	 3.573	&	 -1.474	&	 71.9(6.59)	&	 42.6(5.78)	&	 2  \\
5049.819	&	 &  	 2.278	&	 -1.420	&	 160.4(6.77)	&	 -		&	 1  \\
5109.652	&	 &  	 4.302	&	 -0.980	&	 62.3(6.77)	&	 -		&	 1  \\
5127.359	&	 &  	 0.915	&	 -3.307	&	 133.2(6.37)	&	 -		&	 1  \\
5159.058	&	 &  	 4.283	&	 -0.820	&	 55.3(6.46)	&	 -		&	 1  \\
5187.915	&	 &  	 4.143	&	 -1.260	&	 54.2(6.71)	&	 -		&	 1  \\
5215.179	&	 &  	 3.266	&	 -0.933	&	 121(6.67)	&	 -		&	 1  \\
5242.491	&	 &  	 3.634	&	 -0.840	&	 95.7(6.58)	&	 84.2(5.87)	&	 1  \\
5247.050	&	 &  	 0.087	&	 -4.946	&	     -		&	 136.1(5.89)	&	 1  \\
5250.209	&	 &  	 0.121	&	 -4.938	&	 120.4(6.63)	&	 131.1(5.82)	&	 1  \\
5253.462	&	 &  	 3.283	&	 -1.670	&	 92.1(6.8)	&	 78.6(6.06)	&	 1  \\
5281.790	&	 &  	 3.038	&	 -1.020	&	   -		&	 150.6(5.94)	&	 1  \\
5322.040	&	 &  	 2.280	&	 -2.840	&	 81.2(6.48)	&	 76.4(5.88)	&	 3  \\
5339.930	&	 &  	 3.270	&	 -0.680	&	 132.7(6.63)	&	 -		&	 3  \\
5364.858	&	 &  	 4.445	&	 0.2200	&	 110.5(6.68)	&	 -		&	 3  \\
5365.399	&	 &  	 3.573	&	 -1.440	&	 86.7(6.8)	&	 55.7(5.9)	&	 2  \\
5367.479	&	 &  	 4.415	&	 0.3500	&	 113.5(6.57)	&	 -		&	 1  \\
5369.961	&	 &  	 4.370	&	 0.3500	&	 114.3(6.54)	&	 -		&	 1  \\
5383.369	&	 &  	 4.312	&	 0.5000	&	 130(6.65)	&	 -		&	 1  \\
5543.936	&	 &  	 4.217	&	 -1.140	&	 58.8(6.74)	&	 -		&	 1  \\
5569.620	&	 &  	 3.420	&	 -0.490	&	 125.7(6.45)	&	 -		&	 3  \\
5576.090	&	 &  	 3.430	&	 -0.851	&	 113.5(6.69)	&	 -		&	 1  \\
5586.756	&	 &  	 3.368	&	 -0.210	&	 157.8(6.71)	&	 -		&	 1  \\
5618.631	&	 &  	 4.209	&	 -1.380	&	 51.4(6.72)	&	 -		&	 1  \\
5701.544	&	 &  	 2.559	&	 -2.216	&	 108.5(6.69)	&	 90.7(5.79)	&	 1  \\
5741.848	&	 &  	 4.256	&	 -1.730	&	 20.8(6.63)	&	 -		&	 1  \\
5753.120	&	 &  	 4.260	&	 -0.760	&	 77.6(6.72)	&	 58.6(6.08)	&	 1  \\
5856.088	&	 &  	 4.294	&	 -1.640	&	 34(6.79)	&	 -		&	 1  \\
5859.586	&	 &  	 4.549	&	 -0.386	&	 83.1(6.79)	&	 -		&	 2  \\
5862.357	&	 &  	 4.549	&	 -0.051	&	 81.7(6.31)	&	 -		&	 2  \\
5956.692	&	 &  	 0.859	&	 -4.605	&	 95.9(6.62)	&	 81.6(5.81)	&	 1  \\
6003.010	&	 &  	 3.881	&	 -1.120	&	    -		&	 47.7(5.82)	&	 1  \\
6082.710	&	 &  	 2.222	&	 -3.573	&	 55.2(6.66)	&	 28.3(5.86)	&	 1  \\
6136.994	&	 &  	 2.198	&	 -2.950	&	 84.5(6.47)	&	 78.3(5.88)	&	 1  \\
6137.694	&	 &  	 2.588	&	 -1.403	&	 157.5(6.76)	&	 -		&	 1  \\
6151.620	&	 &  	 2.180	&	 -3.290	&	 79.8(6.71)	&	 62.4(6.01)	&	 3  \\
6173.340	&	 &  	 2.220	&	 -2.880	&	 101.7(6.72)	&	 108.2(6.17)	&	 3  \\
6180.204	&	 &  	 2.727	&	 -2.780	&	 64.8(6.64)	&	 43.2(5.96)	&	 1  \\
6200.314	&	 &  	 2.608	&	 -2.437	&	   -		&	 85(5.96)	&	 1  \\
6213.429	&	 &  	 2.222	&	 -2.660	&	 113(6.68)	&	 102.7(5.88)	&	 1  \\
6219.279	&	 &  	 2.198	&	 -2.433	&	 120.5(6.56)	&	 140.4(6.08)	&	 1  \\
6240.646	&	 &  	 2.222	&	 -3.380	&	 70.8(6.49)	&	 54.1(5.84)	&	 1  \\
6252.554	&	 &  	 2.404	&	 -1.687	&	 155.1(6.73)	&	 152.2(5.76)	&	 1  \\
6254.258	&	 &  	 2.279	&	 -2.480	&	    -		&	 117.3(5.86)	&	 1  \\
\hline
\end{tabular}}

The numbers within the parenthesis in columns 5-6 give the derived 
abundances from the respective line. \\
\end{table*}
}

{\footnotesize
\begin{table*}
{\bf Table A1 }continues... \\ 
\resizebox{\textwidth}{!}{
\begin{tabular}{lccccccc}
\hline                       
Wavelength(\r{A}) & El & $E_{low}$(eV) & log gf & LAMOSTJ091608.81+230734.6 & LAMOSTJ151003.74+305407.3 & Ref  \\ 
\hline 
6280.617	&	 &  	 0.859	&	 -4.390	&	 120.5(6.75)	&	 -		&	 1  \\
6297.800	&	 &  	 2.222	&	 -2.740	&	 90.3(6.36)	&	 -		&	 1  \\
6301.500	&	 &  	 3.654	&	 -0.672	&	 107.8(6.45)	&	 -		&	 2  \\
6322.690	&	 &  	 2.588	&	 -2.426	&	    -		&	 99.7(6.08)	&	 2  \\
6335.328	&	 &  	 2.198	&	 -2.230	&	 128(6.46)	&	 -		&	 1  \\
6393.602	&	 &  	 2.432	&	 -1.620	&	 158.3(6.71)	&	 -		&	 1  \\
6408.016	&	 &  	 3.686	&	 -1.048	&	 93.5(6.52)	&	 -		&	 2  \\
6419.950	&	 &  	 4.730	&	 -0.090	&	 81.3(6.77)	&	 53.5(6.05)	&	 3  \\
6421.349	&	 &  	 2.278	&	 -2.027	&	    -		&	 153.1(5.91)	&	 1  \\
6430.850	&	 &  	 2.180	&	 -2.010	&	 151.5(6.63)	&	 160.5(5.84)	&	 3  \\
6481.870	&	 &  	 2.278	&	 -2.984	&	 99.5(6.8)	&	 -		&	 1  \\
6574.227	&	 &  	 0.990	&	 -5.040	&	 73.2(6.81)	&	 69.5(6.23)	&	 1  \\
6575.019	&	 &  	 2.588	&	 -2.820	&	 79.4(6.69)	&	 -		&	 1  \\
6592.910	&	 &  	 2.720	&	 -1.470	&	 131.8(6.53)	&	 -		&	 3  \\
6593.871	&	 &  	 2.432	&	 -2.422	&	 107.1(6.55)	&	 -		&	 1  \\
6677.989	&	 &  	 2.692	&	 -1.470	&	 148.4(6.64)	&	 169(6.07)	&	 1  \\
6739.521	&	 &  	 1.557	&	 -4.950	&	 28.6(6.59)	&	 -		&	 1  \\
6750.150	&	 &  	 2.424	&	 -2.621	&	 113.7(6.82)	&	 -		&	 1  \\
4416.830	&	Fe II &  2.778	&	 -2.600	&	 -		&	 71.3(5.95)	&	 1  \\
4629.339	&	 &  	 2.807	&	 -2.280	&	 108.3(6.55)	&	 -		&	 1  \\
4923.927	&	 &  	 2.891	&	 -1.320	&	 159.8(6.68)	&	 -		&	 1  \\
5234.620	&	 &  	 3.220	&	 -2.240	&	 99.2(6.48)	&	 60.9(5.76)	&	 3  \\
6247.550	&	 &  	 3.890	&	 -2.340	&	 57.1(6.6)	&	 20.3(5.98)	&	 3  \\
6369.462	&	 &  	 2.891	&	 -4.253	&	 29.7(6.81)	&	 -		&	 2  \\
6456.383	&  &    3.903	&	 -2.075	&	 67.3(6.56)	&	 30.8(6.03)	&	 1    \\	
\hline
\end{tabular}}

The numbers within the parenthesis in columns 5-6 give the derived 
abundances from the respective line. \\
References: 1. F\"uhr et al. (1988), 2. Kurucz (1988), 3. Lambert et al. (1996) \\
\end{table*}
}

{\footnotesize
\begin{table*}
{\bf Table A2 : Equivalent widths (in m\r{A}) of lines used for deriving 
elemental abundances}
\resizebox{\textwidth}{!}{
\begin{tabular}{lccccccc}
\hline                       
Wavelength(\r{A}) & El & $E_{low}$(eV) & log gf & LAMOSTJ091608.81+230734.6 & LAMOSTJ151003.74+305407.3  & Ref  \\ 
\hline 
5682.633     &     Na I     &     2.102     &     -0.700     &     109.4(6.15)     &     -     &     1 \\
5688.205     &          &     2.100     &     -0.450     &     106.8(5.86)     &     -     &     1 \\
6154.226     &          &     2.102     &     -1.560     &     59.90(6.16)     &     23.10(5.27)     &     1  \\
6160.747     &          &     2.104     &     -1.260     &     66.40(5.97)     &     35.20(5.20)     &     1  \\
4702.991     &     Mg I     &     4.346     &     -0.666     &     -     &     	116.9(5.85)     &      2 \\
5528.405     &          &     4.346     &     -0.620     &     175.6(7.23)     &     138.9(6.00)     &     2 \\
5711.088     &          &     4.346     &     -1.833     &     92.80(7.01)     &     64.30(6.20)     &      2 \\
5690.425     &     Si I     &     4.929     &     -1.870     &     31.90(6.67)     &     -     &      3 \\
5948.541     &          &     5.083     &     -1.230     &     61.40(6.72)     &     -     &      3 \\
6145.016     &          &     5.616     &     -0.820     &     31.60(6.38)     &     -     &      4 \\
4435.679     &     Ca I     &     	1.890     &     -0.520     &     -     &     	119.2(4.77)     &   5  \\
5349.465     &          &     2.709     &     -1.178     &     38.20(5.42)     &     -     &     5 \\
5512.980     &          &     2.932     &     -0.290     &     69.90(5.36)     &     -     &     5 \\
5581.965     &          &     2.523     &     -1.833     &     -     &     	39.60(4.41)     &     5 \\
5590.114     &          &     2.521     &     -0.710     &     89.50(5.67)     &     35.30(4.33)     &     5  \\
5857.451     &          &     2.932     &     0.23     &     93.60(5.27)     &     76.20(4.47)     &     5  \\
6102.723     &          &     1.879     &     -0.890     &     102.0(5.26)     &     103.5(4.59)     &     5  \\
6166.439     &          &     2.521     &     -0.900     &     64.10(5.33)     &     -     &     5 \\
6169.042     &          &     2.523     &     -0.550     &     79.50(5.26)     &     -     &     5 \\
6169.563     &          &     2.523     &     -0.270     &     -     &     	83.30(4.53)     &     5  \\
6439.075     &          &     2.525     &     0.47     &     157.5(5.64)     &     -     &     5 \\
6449.808     &          &     2.523     &     -0.550     &     86.90(5.37)     &     70.30(4.61)     &     5 \\
6471.662     &          &     2.525     &     -0.590     &     -     &     	56.80(4.48)     &     5  \\
6493.781     &          &     2.521     &     0.14     &     117.9(5.27)     &     -     &     5 \\
\hline
\end{tabular}}

The numbers within the parenthesis in columns 5-6 give the derived abundances from the respective line. \\
\end{table*}
}

{\footnotesize
\begin{table*}
{\bf Table A2 }continues... \\ 
\resizebox{\textwidth}{!}{
\begin{tabular}{lccccccc}
\hline                       
Wavelength(\r{A}) & El & $E_{low}$(eV) & log gf & LAMOSTJ091608.81+230734.6 & LAMOSTJ151003.74+305407.3  & Ref  \\ 
\hline 
4512.734     &     Ti I     &     0.836     &     -0.480     &     84.90(4.04)     &     -     &      6 \\
4617.269     &          &     1.749     &     0.389     &     87.00(4.29)     &     -     &      6 \\
4759.272     &          &     2.255     &     0.514     &     40.30(3.82)     &     -     &      6 \\
4840.874     &          &     0.899     &     -0.509     &     81.20(3.98)     &     78.00(3.19)     &      6 \\
5007.210     &          &     0.820     &     0.1700     &     131.4(4.25)     &     -               &      6      \\
5024.842     &          &     0.818     &     -0.602     &     82.10(3.95)     &     90.20(3.30)     &      6 \\
5210.386     &          &     0.047     &     -0.884     &     110.0(3.79)     &     143.9(3.23)     &      6 \\
5460.499     &          &     0.050     &     -2.880     &     27.30(4.26)     &     -     &      7 \\
5918.535     &          &     1.070     &     -1.460     &     33.30(4.16)     &     25.70(3.50)     &      6 \\
5922.110     &          &     1.046     &     -1.466     &     35.10(4.17)     &     -     &      6 \\
5941.751     &          &     1.050     &     -1.510     &     30.70(4.14)     &     -     &      6 \\
6303.757     &          &     1.443     &     -1.566     &     25.00(4.03)     &     -     &      6 \\
4568.314     &     Ti II     &     1.220     &     -2.650     &     69.10(3.98)     &     30.20(3.02)     &      6 \\
4571.960     &          &     1.571     &     -0.530     &     -     &     	140.4(3.16)     &      6 \\
4764.526     &          &     1.236     &     -2.770     &     -     &     	49.90(3.47)     &      6 \\
4798.521     &          &     1.080     &     -2.430     &     72.10(3.62)     &     -     &      6 \\
4865.612     &          &     1.116     &     -2.610     &     71.30(3.82)     &     -     &      6 \\
5185.900     &          &     1.890     &     -1.350     &     82.10(3.66)     &     86.30(3.34)     &      6 \\
5381.015     &          &     1.566     &     -2.080     &     -     &     	46.00(3.07)     &     6    \\
5247.565     &     Cr I       &     0.961     &     -1.640     &     -     &     	81.80(3.60)     &      6 \\
5296.691     &          &     0.982     &     -1.400     &     131.8(5.17)     &     -     &     6      \\
5300.744     &          &     0.982     &     -2.120     &     -     &     	49.00(3.68)     &      6 \\
5345.801     &          &     1.003     &     -0.980     &     150.1(5.12)     &     128.7(3.64)     &      6 \\
5348.312     &          &     1.003     &     -1.290     &     121.6(4.86)     &     106.5(3.62)     &      6 \\
5409.772     &          &     1.030     &     -0.720     &     156.1(4.98)     &     130.0(3.41)     &      6 \\
5787.965     &          &     3.323     &     -0.083     &     67.70(5.30)     &     -     &      6 \\
6362.862     &          &     0.941     &     -3.623     &     33.00(5.33)     &     -       &  5      \\
4686.207     &     Ni I     &     3.597     &     -0.640     &     64.70(5.44)     &     -     &      5 \\
4752.415     &          &     3.658     &     -0.700     &     -     &     	40.90(4.76)     &      8 \\
4821.130     &          &     4.153     &     -0.850      &     37.40(5.73)     &     -     &      8 \\
4953.200     &          &     3.74     &     -0.670     &     59.80(5.51)     &     22.00(4.45)     &      5 \\
4980.166     &          &     3.606     &     -0.110     &     -     &     	72.90(4.55)     &      8 \\
5082.350     &          &     	3.657     &     -0.540     &     77.40(5.63)     &     36.50(4.49)     &     8   \\
5102.960     &          &     1.676     &     -2.620     &     -     &     	93.90(4.87)     &      8 \\
6086.280     &          &     4.266     &     -0.530     &     53.50(5.79)     &     -     &      8 \\
6175.360     &          &     4.089     &     -0.530     &     52.30(5.55)     &     -     &      8 \\
6177.236     &          &     1.826     &     -3.500     &     38.40(5.56)     &     -     &      8 \\
6186.710     &          &     4.106     &     -0.777     &     41.80(5.62)     &     -     &      8 \\
6204.600     &          &     4.088     &     -1.130     &     32.40(5.76)     &     -     &      8 \\
6327.593     &          &     1.676     &     -3.150     &     79.10(5.72)     &     -     &      8 \\
6378.247     &          &     4.154     &     -0.890     &     31.40(5.57)     &     -     &     5   \\
6643.629     &          &     1.676     &     -2.300     &     -     &     	106.6(4.53)     &     1 \\
4722.150     &     Zn I     &     4.029     &     -0.370     &     -     &     	34.80(2.66)     &      9 \\
4810.530     &          &     4.080     &     -0.170     &     -     &     	33.80(2.46)     &      9 \\
6362.338     &          &     0.150     &     5.796     &     29.60(4.11)     &     -     &      10 \\ 
4607.327     &     Sr I     &     0.000     &     -0.570     &     84.20(3.21)     &     -     &      11 \\
6435.004     &     Y I      &     0.066     &     -0.820     &     47.70(2.73)     &     55.20(2.19)     &      12 \\
4883.684     &     Y II     &     1.084     &     0.07     &     145.4(2.37)     &     -     &      12 \\
5119.112     &          &     0.992     &     -1.360     &     -     &     	172.7(3.13)     &      12 \\
5289.815     &          &     1.033     &     -1.850     &     63.50(2.27)     &     -     &      12 \\
5402.774     &          &     1.839     &     -0.510     &     77.50(2.20)     &     -     & 13  \\
5544.611     &          &     1.738     &     -1.090     &     64.50(2.36)     &     -     &      12 \\
5546.009     &          &     1.748     &     -1.110     &     -     &     	124.9(2.97)     &      12 \\
5662.925     &          &     1.944     &     0.16     &     105.3(2.26)     &     -     &      13 \\
6613.733     &          &     	1.748       &     -1.100     &     74.20(2.49)     &     -     &      12 \\
\hline
\end{tabular}}

The numbers within the parenthesis in columns 5-6 give the derived abundances from the respective line. \\
\end{table*}
}

{\footnotesize
\begin{table*}
{\bf Table A2 }continues... \\ 
\resizebox{\textwidth}{!}{
\begin{tabular}{lccccccc}
\hline                       
Wavelength(\r{A}) & El & $E_{low}$(eV) & log gf & LAMOSTJ091608.81+230734.6 & LAMOSTJ151003.74+305407.3  & Ref  \\ 
\hline
4739.480     &     Zr I     &     0.651     &     0.230     &     84.30(3.30)     &     47.30(1.89)     &      14 \\
4772.323     &          &     0.623     &     0.044     &     60.90(2.92)     &     63.90(2.27)     &      14 \\
4805.889     &          &     0.687     &     -0.420     &     40.40(3.06)     &     -     &      14 \\
6134.585     &          &     0.000     &     -1.280     &     54.10(3.18)     &     32.50(2.15)     &      14 \\
4257.120     &     Ce II     &     0.460     &     -1.116     &     -     &     	36.80(1.15)     &      15 \\
4336.244     &          &     	0.704     &     -0.564     &     82.70(2.32)     &     -     &     15   \\
4349.789     &          &     	0.701     &     -0.107     &     -     &     	65.90(1.34)     &     13 \\
4407.273     &          &     0.701     &     -0.741     &     -     &     	53.70(1.32)     &      15 \\
4497.846     &          &     0.958     &     -0.349     &     -     &     	67.00(1.44)      &    15    \\
4508.079     &          &     	0.621     &     -1.238     &     60.90(2.28)     &     -     &      15 \\
4628.161     &          &     0.516     &     0.008     &     120.8(2.43)     &     -     &      15 \\
4747.167     &          &     0.320     &     -1.246     &     -     &     	50.80(1.25)     &      15 \\
4873.999     &          &     1.107     &     -0.892     &     49.60(2.22)     &     -     &      15 \\
5187.458     &          &     1.211     &     -0.104     &     86.20(2.35)     &     -     &     	15  \\
5274.229     &          &     1.044     &     -0.323     &     -     &     	96.10(1.21)     &      6 \\
5330.556     &          &     0.869     &     -0.760     &     72.30(2.23)     &     59.40(1.53)     &      15 \\
6034.205     &          &     1.458     &     -1.019     &     42.10(2.52)     &     -     &      15 \\
5188.217     &     Pr II     &     0.922     &     -1.145     &     26.00(1.59)     &     20.40(1.18)     &      15 \\
5219.045     &          &     	0.795     &     -0.240     &     -     &     	77.20(1.02)     &      16 \\
5259.728     &          &     0.633     &     -0.682     &     65.90(1.60)     &     61.90(1.04)     &      16 \\
5292.619     &          &     0.648     &     -0.300     &     83.40(1.62)     &     -     &      16 \\
5322.772     &          &     0.482     &     -0.315     &     90.30(1.58)     &     98.00(0.98)     &      15 \\
6165.891     &          &     0.923     &     -0.205     &     59.20(1.24)     &     63.00(0.87)     &      15 \\
6278.676     &          &     1.196     &     -0.630     &     22.20(1.23)     &     -     &      16 \\
4446.384     &     Nd II & 	0.204     &     0.590     &     -     &     	110.9(1.27)     &      17 \\
4451.563     &          &     0.380     &     -0.040     &     115.1(2.04)     &     -     &      17 \\
4556.133     &          &     0.064     &     -1.610     &     76.60(2.13)     &     -     &      15 \\
4811.342     &          &     0.064     &     -1.140     &     92.80(2.00)     &     -     &      15 \\
4825.478     &          &     0.182     &     -0.860     &     112.4(2.37)     &     -     &      15 \\
4947.020     &          &     0.559     &     -1.250     &     -     &     	58.00(1.42)     &      15 \\
4961.387     &          &     0.631     &     -0.710     &     96.00(2.32)     &     98.20(1.56)     &      15 \\
5212.361     &          &     0.204     &     -0.870     &     -     &     	104.7(1.21)     &      15 \\
5276.869     &          &     0.859     &     -0.440     &     -     &     	71.80(1.15)     &      17 \\
5287.133     &          &     0.744     &     -1.300     &     -     &     	39.30(1.39)     &      15 \\
5356.967     &          &     1.264     &     -0.250     &     -     &     	68.20(1.42)     &      17 \\
5361.510     &          &     0.680     &     -0.400     &     -     &     	92.40(1.16)     &        17 \\
5442.264     &          &     0.680     &     -0.910     &     84.40(2.19)     &     86.00(1.56)     &      17 \\
5485.696     &          &     1.264     &     -0.120     &     77.70(1.96)     &     68.70(1.28)     &      17 \\
5603.648     &          &     0.380     &     -1.830     &     71.70(2.43)     &     -     &      15 \\
5718.118     &          &     1.410     &     -0.340     &     -     &     	48.90(1.39)     &      17 \\
5825.857     &          &     1.080     &     -0.760     &     65.90(2.08)     &     50.00(1.40)     &      15 \\
4458.509     &     Sm II     &     0.104     &     -1.110     &     76.30(1.45)     &     -     &      15 \\
4499.475     &          &     0.248     &     -1.413     &     52.10(1.34)     &     61.20(1.04)     &      15 \\
4519.630     &          &     0.543     &     -0.751     &     -     &     	81.70(1.07)     &      15 \\
4566.210     &          &     0.330     &     -1.245     &     57.80(1.38)     &     63.70(1.01)     &        15 \\
4615.444     &          &     0.544     &     -1.262     &     50.60(1.49)     &     -     &      15 \\
4642.228     &          &     0.379     &     -0.951     &     71.20(1.45)     &     88.90(1.15)     &      15 \\
4674.593     &          &     0.184     &     -1.055     &     68.10(1.23)     &     -     &      15 \\
4676.902     &          &     0.040     &     -1.407     &     62.30(1.28)     &     -     &      15 \\
4704.400     &          &     0.000     &     -1.562     &     73.10(1.62)     &     65.00(0.89)     &     15   \\
4726.026     &          &     0.333     &     -1.849     &     31.80(1.42)     &     27.70(1.02)  & 15  \\
4854.368     &          &     	0.379     &     -1.873     &     32.10(1.49)     &     -     &      15 \\
\hline
\end{tabular}}

The numbers within the parenthesis in columns 5-6 give the derived abundances from the respective line. \\
References: 1. Kurucz et al. (1975), 2. Lincke et al. (1971), 3. Garz (1973),
4. Schulz-Gulde (1969), 5. Kurucz (1988), 6. Martin et al. (1988), 7. Smith et al. (1978), 8. F\"uhr et al. (1988), 
9. Warner (1968), 10. Lambert et al. (1969), 11. Corliss et al. (1962), 12. Hannaford et al. (1982), 13. Cowley et al. (1983), 14. Biemont et al. (1981), 
15. Meggers et al. (1975), 16. Lagehaling et al. (1976), 17. Ward et al. (1985) \\
\end{table*}
}

\end{document}